\documentclass{aastex631}

\usepackage{todonotes}

\usepackage{ulem}
\let\tablenum\relax
\usepackage{siunitx}
\usepackage{enumitem}

\shorttitle{SNAPS}
\shortauthors{Trilling et al.}
\graphicspath{{./}{figures/}}

\begin{document}

\title{The Solar System Notification Alert Processing System (SNAPS): Design, Architecture, and First Data Release (SNAPShot1)}

\author[0000-0003-4580-3790]{David E. Trilling}
\affiliation{Department of Astronomy and Planetary Science \\
Northern Arizona University \\
P.O. Box 6010 \\
Flagstaff, AZ 86011, USA}
\affiliation{School of Informatics, Computing, and Cyber Systems \\
Northern Arizona University \\
P.O. Box 5693 \\
Flagstaff, AZ 86011, USA}

\author[0000-0002-0826-6204]{Michael Gowanlock}
\affiliation{School of Informatics, Computing, and Cyber Systems \\
Northern Arizona University \\
P.O. Box 5693 \\
Flagstaff, AZ 86011, USA}
\affiliation{Department of Astronomy and Planetary Science \\
Northern Arizona University \\
P.O. Box 6010 \\
Flagstaff, AZ 86011, USA}

\author[0000-0002-6676-1713]{Daniel Kramer}
\affiliation{School of Informatics, Computing, and Cyber Systems \\
Northern Arizona University \\
P.O. Box 5693 \\
Flagstaff, AZ 86011, USA}
\affiliation{Department of Astronomy and Planetary Science \\
Northern Arizona University \\
P.O. Box 6010 \\
Flagstaff, AZ 86011, USA}

\author{Andrew McNeill}
\affiliation{Department of Astronomy and Planetary Science \\
Northern Arizona University \\
P.O. Box 6010 \\
Flagstaff, AZ 86011, USA}
\affiliation{Department of Physics \\ 
Lehigh University \\
16 Memorial Drive East \\
Bethlehem, PA 18015, USA}

\author{Brian Donnelly}
\affiliation{School of Informatics, Computing, and Cyber Systems \\
Northern Arizona University \\
P.O. Box 5693 \\
Flagstaff, AZ 86011, USA}

\author{Nat Butler}
\affiliation{School of Earth and Space Exploration \\
Arizona State University \\ 
Tempe, AZ 85287, USA}

\author[0000-0003-1204-6535]{John Kececioglu}
\affiliation{Department of Computer Science \\
University of Arizona \\ 
Tucson, AZ 85721}



\begin{abstract}
We present here the design, architecture, and first data release for the Solar System Notification Alert Processing System (SNAPS). SNAPS is a Solar System broker that ingests alert data from all-sky surveys. At present, we ingest data from the Zwicky Transient Facility (ZTF) public survey, and we will ingest data from the forthcoming Legacy Survey of Space and Time (LSST) when it comes online. SNAPS is an official LSST downstream broker. In this paper we present the SNAPS design goals and requirements. We describe the details of our automatic pipeline processing in which physical properties of asteroids are derived. We present SNAPShot1, our first data release, which contains 5,458,459~observations of
31,693~asteroids observed by ZTF
from July, 2018, through May, 2020.
By comparing a number of derived properties for this ensemble to previously published results for overlapping objects we show that our automatic processing is highly reliable. We present a short list of science results, among many that will be enabled by our SNAPS catalog: (1) we demonstrate that there are no known asteroids with very short periods and high amplitudes, which clearly indicates that in general asteroids in the size range 0.3--20~km are strengthless; (2) we find no difference in the period distributions of Jupiter Trojan asteroids, implying that the L4 and L5 cloud have different shape distributions; and (3) we highlight several individual asteroids of interest.
Finally, we describe future work for SNAPS and our ability to operate at LSST scale.
\end{abstract}



\section{Introduction}

\subsection{Asteroid science from sky surveys}

Asteroids are the most numerous objects in the Solar System and as
such act as tracers of the dynamical and physical evolution of our
planetary system. At present, there are more than 1~million known
asteroids.
A comprehensive understanding of the
evolution of our Solar System can therefore use the individual
properties of asteroids as the fine-grained measurements from which
broader conclusions can be drawn.
Asteroids can be used to understand
the dynamical and collisional history of the Solar System,
the intrinsic material properties of primitive bodies,
and help unravel the origin of life on Earth through understanding
the degree to which asteroids brought water and organics to the early
Earth.
Individual asteroids are akin to
pixels in a large image, and by building up our knowledge, one pixel
at a time, the entire picture is revealed.

To gain this global understanding of the Solar System through asteroids, large
datasets are required.
The Vera C. Rubin Observatory will carry out the {\em
  Legacy Survey of Space and Time (LSST)} starting in 2024 and
will
revolutionize many fields of astronomy.
LSST will observe more than 5 million main belt asteroids \citep{lsstbook}.
It is not only the number of asteroids to be observed that is significant, but the fact that most asteroids will be observed many hundreds of times, under relatively uniform conditions. This data set will be far richer --- and far more scientifically
valuable --- than existing catalogs of asteroid measurements.
However,
the LSST data by itself (time, brightness, position) will not provide
new insight into the processes that have driven the evolution of
our Solar System.
To reap the scientific reward from this data set, the
scientific context and physical properties of the entire asteroid population must be
understood.

The current LSST schedule of first light in 2023 and science operations in 2024 means that the next steps in understanding the history of the Solar System are not far away. 
This has motivated us to develop the necessary infrastructure now, to be prepared for the arrival of LSST data. 
Fortunately, the ongoing {\em Zwicky Transient Facility} (ZTF; \citealt{ztf}) survey is a
present-day analog of and precursor to LSST. We are using data from ZTF
to develop, hone, and prepare for the coming LSST era.

All-sky surveys enable two important research goals:
(1) understanding global trends, and (2) looking for rare events, as follows.
In category~1, by deriving physical properties of millions of asteroids
we can understand the global properties of the asteroid belt that
reveal information about the formation and evolution of our
Solar System. These are signatures that are only revealed through very large sample sizes. Examples here could include measuring asteroid lightcurve periods and amplitudes as a function of color (and implied composition), where
the required densities and strengths may differ between
rocky S-type asteroids and more primitive C-type asteroids. Large data sets can enable detecting subtle signatures, or slicing datasets (example: as a function of size) to reveal trends.

Objects detected in category~2 --- 
detecting and characterizing
rare events --- also can place constraints on Solar System formation through probing subtle signatures that can indicate important processes.
We refer to objects that exhibit rare behaviors as outliers, and there are two general kinds. Real-time
  outliers (category~2a) are objects whose properties change on
short timescales.
An example of this case is an asteroid
that is found to be active.
Population outliers (category~2b) are
objects whose fundamental properties are unusual compared to the
entire population. Examples here could include asteroids with very
large lightcurve amplitudes; very short or long rotation periods; or
asteroids with very unusual colors.

Studying outlier asteroids allows us to understand these
rare events. As an example in category~2a, it is generally thought that most activity
among
asteroids is driven by sublimation of volatiles, but
the
occurrence
rate of active asteroids is not well known
(e.g., \citealt{2015aste.book..221J}).
Measuring this rate from a uniform survey would improve
our understanding of the volatile content of the asteroid
belt and provide important new evidence for understanding the
formation of the Solar System and the origin of life on Earth.
(The LOOK project \citep{look} has some goals and approaches that overlap with SNAPS, though LOOK has an overall greater emphasis on cometary science.)
Thus,
real-time outlier detections provide direct impact on these
important outstanding science questions.
Population outliers (category~2b) may also provide insight into the
formation
and history of the Solar System. For example, a small number of
asteroids have very high amplitude lightcurves or very long rotation
periods,
both of which may indicate the need for significant internal strength
(e.g., 
\citealt{2018AJ....156..282M}
and references therein).
This too has implications for how asteroids formed, and how the Solar
System
formed.

\subsection{ZTF and LSST}

ZTF does, and LSST will, broadcast {\em alerts}
that report transient objects on the sky.
These variable sources include supernovae, variable stars, and moving objects (and many other kinds of astrophysically variable objects).
Moving objects are considered variable here because the brightness at a given sky location changes as a function of time, as an asteroid passes to and from that location. 
An alert includes measured data,
some properties of the observed source, a postage stamp, and metadata, and alerts are distributed through a Kafka stream.
LSST will
broadcast {\em alerts} for around 10,000~variable sources detected in each 30~second visit\footnote{\url{ https://dmtn-102.lsst.io/DMTN-102.pdf}}. Moving objects may be 10\% to 50\% of all variable sources in a given LSST visit, depending primarily on ecliptic latitude.
The ZTF data flow is about one tenth of this rate, with some 1000~alerts, and 100~asteroids, per visit, which corresponds to around 10,000~asteroids
each night.
The alerts (will) include
measured properties of moving objects, including
date/time; photometric properties (magnitude, uncertainty);
a real/bogus score (likelihood that the source is real and is a point source);
and properties of the PSF (e.g., elongation or extendedness).
In the work described here, we focus only on {\em known objects}; in this
case,
the identity of the object is also transmitted with the measurement
data. The orbital elements of each reported object are therefore
known; in the work presented here, the orbital elements are not transmitted with the measurements and we obtain them through cross-referencing the Minor Planet
Center's catalogs.

At full scale, LSST will report measurements of
$10^5$ -- $10^6$~asteroids
every night, for ten years. The total
observational record, over ten years, will be a few
billion unique measurements of more than
5~million unique asteroids. 
LSST is a carefully defined project,
and 
key definitions for the LSST project --- what LSST will and will not
do ---
are clearly defined
(see
\citealt{lsstsciencedrivers}
and many documents at the 
LSST Documentation Hub\footnote{\url{https://www.lsst.io/}}).
This data stream will enable science, but will not produce science.
In order to carry out a wide range of deep scientific
investigations, the detailed physical properties
of these millions of asteroids must be derived outside
of the LSST project.

This flood of data can only be handled with sophisticated automated software. The general approach to handling Solar System objects, or indeed any transient or time-varying science, is through {\em brokers}: software packages that automatically ingest the LSST (or ZTF) alert stream. Brokers have some intelligence and may have the ability to derive target properties, recognize outlier behavior, attach external measurements to create value-added data records, and issue further alerts. There are a number of brokers that are under development and that are oriented towards sidereal or astrophysical sources, or are general and will include some basic processing of Solar System observations\footnote{\url{https://www.lsst.org/scientists/alert-brokers}}. 

Individual asteroid measurements such as magnitude or position by themselves are generally not
scientifically interesting (except for extendedness, as described
below).
However, for asteroids with enough measurements, additional properties can be derived, including
Solar System absolute magnitude ($H$); photometric slope parameter
($G$);
lightcurve period and amplitude;
color; taxonomic classification, as estimated from color; and
even albedo and diameter, as estimated from color \citep{ivezicivezic}.

\subsection{This paper: SNAPS}

In this paper we present SNAPS --- the Solar System Notification Alert Processing System. SNAPS was designed from the outset to be a moving object broker and hence has a number of capabilities that specifically enable Solar System science. SNAPS has been designated an official downstream broker.
Here we present, in Section~\ref{sec:design}, the SNAPS design and architecture and our technical work completed to date.
In Section~\ref{sec:snapshot1} we present SNAPShot1, our first data release.
In Section~\ref{sec:accuracy} we demonstrate the accuracy of our automatically derived results.
Section~\ref{sec:science} presents several science results, among the many that will be enabled by our SNAPS catalog.
In Section~\ref{sec:discussion} we discuss the scalability of our approach to LSST operations, and Section~\ref{sec:future} describes the next steps for SNAPS.

\section{SNAPS design \label{sec:design}}

The SNAPS broker has been designed to ingest the ZTF alert stream and enable science, as well as act as a testbed for broker needs for the LSST era. Over the past four years, since the beginning of the ZTF real-time alert broadcast in July, 2018, we have ingested nearly 20~million observations of nearly 600,000~unique asteroids. In this section we present
various elements of the overall design of SNAPS.

%

\subsection{Overview}

\begin{figure}
    \begin{center}
    \includegraphics[width=0.45\textwidth]{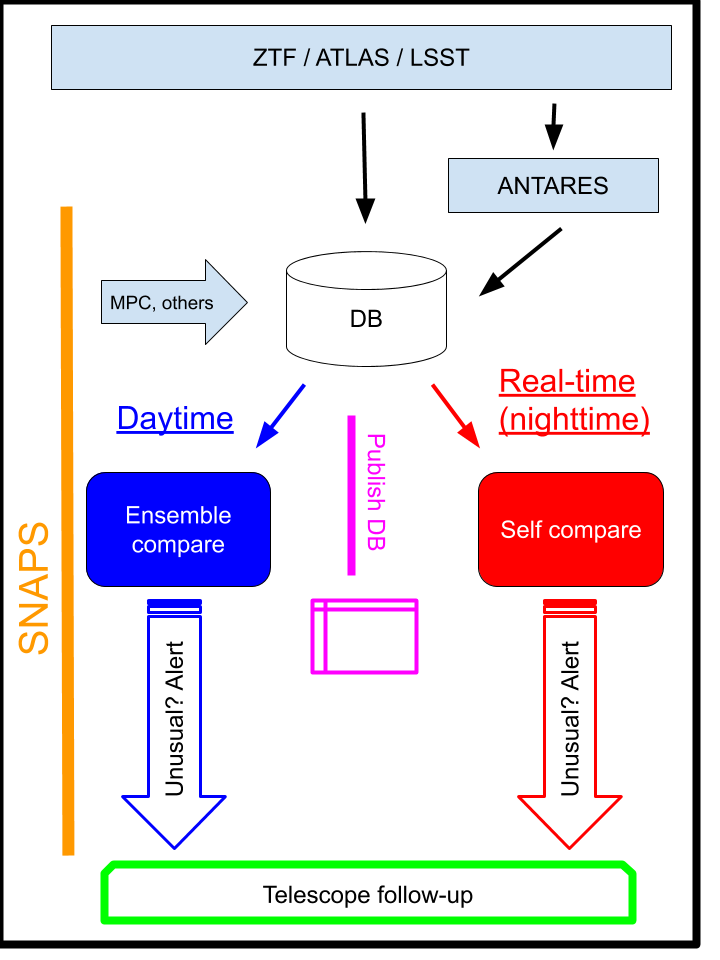}
    \end{center}
    \caption{
      Flowchart for
      SNAPS: the {\em Solar System Notification Alert Processing System}.
      The scope of the overall SNAPS project is indicated by the orange bar
  on the side.
  The main SNAPS facets are (a) ingesting data, from a survey or from an upstream broker; (b) detecting real-time outliers [red path here]; (c) detecting population outliers [blue path here]; and (d) disseminating results. Telescope follow-up may be appropriate but is not formally part of SNAPS.}
    \label{fig:flowchart}
\end{figure}

\noindent The high-level description of SNAPS is that it must ingest data from an alert stream; update the database with this recent data; search for both real-time and population outliers; and disseminate results through both an alert stream and a publicly accessible web portal.
The architecture for
SNAPS is shown in
Figure~\ref{fig:flowchart}.
Our broker has four primary facets:
(a) We listen to an alert stream, either directly from the
source (ZTF/LSST), or through an intermediate broker
(in this case, ANTARES;
\citealt{antares}).
We also incorporate adjacent databases
(e.g, MPC, SDSS, NEOWISE, 2MASS, etc.).
(b) We analyze in {\em real time}
new observations, and
issue alerts if unusual properties
are detected (red path in Figure~\ref{fig:flowchart}).
Typically this done by comparing the object's observed behavior
with its expected behavior.
(c) During the {\em day time} each object is compared to the
ensemble, and alerts are triggered about
objects with unusual properties.
(d) A {\em value-added database} with derived
properties for all asteroids is published and
world-accessible. This database enables a wide range of science investigations.

Element~(a) is fully mature and is presented here. 
Element~(b) is under development, and will be reported in a future paper.
Element~(c) is presently carried out, though in a manual, rather than automated or algorithmic, way. Element~(d) is partially complete: the value-added database exists, and the public database portal and alert stream are under development.

\subsection{Description of SNAPS elements}

\subsubsection{Element (a): Ingesting data}

ZTF broadcasts an alert stream\footnote{\url{https://www.ztf.caltech.edu/ztf-alert-stream.html}}
with measurements and metadata for all transient sources detected in an image. Around 10\% of all transient sources are Solar System objects. We could listen directly to this alert stream, and discard the 90\% of alerts that are irrelevant for us, but we choose a different approach instead. The ANTARES broker \citep{antares} is broadly interested in all astrophysical transients, though not Solar System objects. ANTARES listens to the full ZTF alert stream. As part of their processing, they rebroadcast Solar System alerts on a dedicated Kafka stream. It is this stream that we ingest.

One of the first steps that ANTARES carries out, before rebroadcasting Solar System objects, is removing only observations with real bogus scores $<0.55$. ZTF defines its real/bogus scores in the range $[0,1]$ where 0~is not a point source and 1~is a point source \citep{realbogus}. 
Since the ZTF real bogus classifier was trained on point sources, real extended objects can receive low real/bogus scores and would not be distributed by ANTARES. To ingest these ``lost" observations, we download and ingest the data from ZTF's alert archive\footnote{\url{https://ztf.uw.edu/alerts/public/}}.
For the data presented in this paper, however, we are not searching for activity, and the ANTARES re-broadcast is sufficient.

Each individual measurement has a large number of metadata properties associated with it; the relevant ones that we store in our database are shown in Table~\ref{tab:ztf_cols1}.

\begin{table}[t]
    \centering
\begin{tabular}{|ll|} \hline
Parameter & Definition \\ \hline
Image information & \\ \hline 
{\tt jd} & {\em Julian date}\\
{\tt fid} & {\em Filter ID (a numberical value showing g,r,i)} \\
{\tt ra} & {\em Right ascension} \\
{\tt dec} & {\em Declination} \\
{\tt night} & {\em Date (YYYMMDD) of observation}  \\
\hline 
Source information & \\ \hline 
{\tt magpsf} & {\em Magnitude from PSF fitting} \\
{\tt sigmagpsf} & {\em 1$\sigma$ uncertainty in magpsf} \\
{\tt chipsf} & {\em Reduced $\chi^2$ for PSF fit} \\
{\tt magap} & {\em Magnitude in 8~pixel aperture} \\
{\tt sigmagap} & {\em 1$\sigma$ uncertainty in magap} \\
{\tt magapbig} & {\em Magnitude in 18~pixel aperture} \\
{\tt sigmagapbig} & {\em 1$\sigma$ uncertainty in magapbig} \\
{\tt elong} & {\em Ratio of semi-major to semi-minor axis of the source} \\
{\tt rb} & {\em Real-bogus score [0 bad, 1 good] -- likelihood of true point source} \\
{\tt ssnamenr} & {\em Name of nearest known Solar System object} \\
{\tt id} & {\em ZTF identifier for this observation} \\
\hline
Values from Horizons & \\ \hline
{\tt obsdist} & {\em Distance from source to observer} \\
{\tt heliodist} & {\em Distance from source to Sun} \\
{\tt phaseangle} & {\em Observer-source-Sun angle} \\
{\tt ltc} & {\em Light time correction} -- calculated\\ \hline
\end{tabular}
\caption{Properties of individual observations that are stored in the SNAPS database.
Image and source information is extracted from the ZTF (or LSST) alert;
the descriptions for the ZTF columns are available at \url{https://zwickytransientfacility.github.io/ztf-avro-alert/schema.html} .
Using {\tt ssnamenr} we retrieve relevant geometry information from JPL's Horizons service.
}
    \label{tab:ztf_cols1}
\end{table}

\begin{table}[t]
    \centering
\begin{tabular}{|ll|} \hline
Parameter & Definition \\ \hline
Derived properties  & \\ \hline
{\tt H} & {\em Solar System absolute magnitude in {\tt g} and {\tt r}, with uncertainties} \\
{\tt G} & {\em Solar System phase parameter in {\tt g} and {\tt r}, with uncertainties} \\
{\tt g-r color} & {\em mean color and uncertainty} \\
{\tt lcper} & {\em Derived lightcurve period} \\
{\tt lcamp} & {\em Derived lightcurve amplitude} \\
{\tt peakpower} & {\em Peak power in the periodogram} \\
{\tt residPeriod} & {\em The derived LS period for the residuals of the generated light curve model}\\
{\tt mag18/mag8} & {\em The max, mean, standard deviation, and most recent mag18/mag8 value for the object}  \\
\hline
Information from external catalogs &  \\ \hline
{\tt ugriz photometry} & {\em SDSS} \\
{\tt JHK photometry} & {\em 2MASS} \\
{\tt diameter and albedo} & {\em WISE/NEOWISE} \\
{\tt orbital elements; H, G} & {\em Minor Planet Center} \\
{\tt Rotation period and quality code} & {\em Lightcurve database} \\ \hline
\end{tabular}
\caption{Derived properties of individual objects and values from external catalogs that are stored in the SNAPS database.
}
    \label{tab:ztf_cols2}
\end{table}

\subsubsection{Element (b): Real-time analysis}

After ingesting the data, we retrieve object ephemerides from JPL Horizons\footnote{\url{https://ssd.jpl.nasa.gov/horizons/}}
by using 
the \texttt{jplhorizons} module in astroquery
\citep{astroquery}.
Using geocentric and heliocentric distance and phase angle we calculate the object's absolute magnitude in the Bowell HG system \citep{1989aste.conf..524B} using 
either 
a previously derived $G$ value if one exists, or else by assuming $G=0.15$.
Light time corrections are also retrieved, to be used in our downstream analysis.
These values are stored in our database Table~\ref{tab:ztf_cols1}.

One goal of our real-time analysis is to compare a new observation with the prediction for that observation and to identify significant differences. The measured properties to be considered here include magnitude and PSF shape. This element of SNAPS is not fully mature at this time.

To calculate predicted magnitude, we will use our derived Solar System absolute magnitude ($H$) and slope parameter ($G$) and the observational geometry to define the expected magnitude, which is then modulated by the object's lightcurve (described in the next section). Clearly, our predictions are more accurate for objects with a large number of observations, where all of $H$, $G$, and lightcurve are well determined.

The vast majority of objects have PSF shapes that are round; this corresponds to ZTF {\tt elong} values of~1.0.
Therefore, an observation with an {\tt elong} value that differs significantly from unity may be of interest. It is also possible that an object that is known to be active has significant changes in its {\tt elong} values, indicating either a significant increase or decrease in activity. In principle, these changes to PSF shape can be determined with a very small number of observations in our database, and the algorithmic implementation is relatively straightforward and is under way.

As described above, 
the ZTF real-bogus algorithm was trained on point sources.
Therefore, asteroids that show signs of activity --- in other words, objects that are expected to be point sources but are not --- may receive low {\tt rb} scores,
and these anomalous observations will also be investigated.

\subsubsection{Element (c): Day-time analysis}

Day-time analysis is designed to enable science results that are not highly time critical. We (will) produce results both for individual objects and for the entire population. \\

\noindent {\bf Individual object processing.}
Objects with more than 50~observations are processed with two separate pipelines to calculate
lightcurve properties and phase curves. (We do not attempt to derive properties for objects with fewer than 50~observations.)
The lightcurve pipeline is used to derive an object's color and lightcurve properties (period and amplitude). The phase curve calculation pipeline is used to derive
absolute magnitude and phase parameter. \\

\noindent {\em Deriving lightcurves.} Lightcurves are calculated in the following way.
\begin{enumerate}
    \item Observations are split into groups by filter.
    \item Each filter group is processed through a Lomb-Scargle Periodogram (LSP) approach \citep{1976Ap&SS..39..447L,1982ApJ...263..835S} using a uniform period grid of $10^6$
    periods, ranging from 1--2500~hours. These values are {\em lightcurve periods}; since an asteroid typically executes two lightcurve periods in a single rotation period, the {\em rotation period} search range is 2--5000~hours.
(It would be very interesting to identify asteroids with rotation periods less than 2~hours -- so-called super fast rotators (SFRs), which are unusual and generally require non-zero strength to avoid catastrophic disruption.
However, the sparse ZTF cadence is not well suited to finding SFRs, and our synthetic population studies, described below, show many false positives rotation periods less than 2~hours. We therefore use a lower limit of 2~hours in this paper, and defer the more complex topic of searching for SFRs in ZTF/SNAPS to a subsequent paper.)

    \item The resulting periods derived separately for each filter are compared. If they are the same, that period is then used as the object's period. If they are not the same, then the filter with the most observations (for ZTF, this is usually $r$) is used as the object's period.
    \item A Fourier model is then fit to the $g$ and $r$ data separately using the derived period. Using the median of each Fourier model, a $g-r$ color can be derived. 
    \item Using the derived color, all data is combined (that is, the $g$ data is shifted using the derived $g-r$ color to overlay it on the unshifted $r$ band data). This combined data is used to derive a refined period solution.
    \item Another Fourier model (LSP) is fit for the combined multi-band data and the amplitude of this solution is treated as the amplitude of the object's lightcurve. 
    \item The derived period, amplitude, Fourier parameters, and $g-r$ color are then stored in our database (Table~\ref{tab:ztf_cols2}).
\end{enumerate}

All ground-based surveys are subject to aliasing in period searches due to the unavoidable 24~hour day/night cycling of potential observing times. Nearly 20\% of our lightcurve period solutions have values in the ranges 11--13~hours,
15--17~hours, 23--25~hours, or 47--49~hours, with the largest contribution from periods near 16~hours. That the dominant alias is at 16~hours, and not 12~or 24~hours, is likely due to details of the ZTF cadence.
Our analysis of these alias features, and our overall
approach to de-aliasing, 
will be presented in a forthcoming paper (Kramer et al.); for this data release, no attempt at de-aliasing is made. Users are cautioned that a substantial fraction of solutions near these common periods (12, 16, 24, 48~hours) are likely to be aliases and not the true rotation periods,
although certainly there are some asteroids whose true periods are near these values. At present, it is impossible to distinguish (false) alias solutions from (true) rotation periods near these common periods
(12, 16, 24, 48~hours).
%
Nevertheless, for all of these objects
the $g-r$ color and amplitude 
are most likely correct. \\

\noindent {\em Phase curve Calculations.} We carry out the following steps to derive phase curve solutions for each object.
Steps~1 and~2 are
carried out separately for data in each filter.

\begin{enumerate}[ref=~\arabic*]
    \item \label{list:pcc:ms} 
    Our initial solution for each asteroid is $H=10$ and $G=0.15$. From these values we iteratively adjust and derive best-fit $H,G$ values using SciPy's minimize function, $1-r_P^2$, where $r_P$ is the Pearson correlation coefficient \citep{freedman2007statistics}. The search bounds for $G$ are ($-0.302$, $0.907$), which are 10\% (less than, greater than) the (minimum, maximum) values from \citet{Vere__2015}.
(The minimization process also requires bounds on $H$; we use [4.7, 29.3], well beyond the expected values for any SNAPS targets.)
    The minimize function returns, among other values, $G$, $H$, and the success of the minimization.
    \item \label{list:pcc:err} Step~1 is repeated 30~times. For each of these trials, the magnitudes for each observation are randomized with a Gaussian distribution within the reported observational errors. The standard deviation of the results from these 30~trials are used as $\sigma_H$ and $\sigma_G$. 
    \item If minimizations in both filters succeed, then the two derived $G$ values are combined, using an average weighted by the number of observations in each filter. This step does not add much in the case of ZTF with only two filters but could be useful for other data sets (i.e., LSST) where solutions exist in three or more filters.
    \item The derived $G$, $H$, and $\sigma$ values and the success of each minimization are stored in the database.
\end{enumerate}


\noindent {\bf Population studies.}
There are very few examples of outlier asteroids in the literature. Consequently, we use an unsupervised approach to detect asteroids that are outliers relative to the population of objects in our ZTF database. We create a feature vector for each asteroid in our database using 15~properties selected from Tables~\ref{tab:ztf_cols1}~and~\ref{tab:ztf_cols2}, where example feature vectors are given in Figure~\ref{fig:example_fvs}.
We use these features as input to several unsupervised outlier detection algorithms. At present, each object is assigned an outlier ranking, $r_i \in[1,N_{objects}]$ (here, $N_{objects}=31693$).
Object $i$ with $r_i=1$ refers to the object with the greatest outlier score, whereas the object with $r_i=31693$ is denoted as the most typical object in the database.

\begin{table}
\centering
\begin{tabular}{|c|r|r|r|r|r|c|r|}
\hline
Object&\texttt{lcamp}&\texttt{lcper}&\texttt{g-r color}&\texttt{peakpower}&\texttt{diameter}&$\ldots$&\texttt{albedo}\\
 & (mag) & (hr) & (mag) & () & (km) &  & \\\hline
152&0.229&6.245&0.679&0.695&57.8&&0.247\\
168&0.161&46.973&0.472&0.595&144.1&&0.045\\
198&0.151&8.532&0.641&0.428&50.9&&0.293\\
227&0.165&15.969&0.473&0.343&109.3&&0.046\\
\hline
\end{tabular}
\caption{Four example feature vectors in SNAPShot1, where a subset of 15 features from Tables~\ref{tab:ztf_cols1}~and~\ref{tab:ztf_cols2} are shown here.} \label{fig:example_fvs}
\end{table}

SNAPS currently ranks each object using two outlier detection methods, although we will expand on the number of outlier detection methods in the future. Our aim is to use an ensemble of methods for outlier detection, as each method will derive disparate sets of outliers, as they are sensitive to different aspects of the data. We combine the output of these methods to derive an average outlier score for each object. 

We use $k$-nearest neighbors ($k$NN) as an outlier detection method~\citep{hautamaki2004outlier}. Intuitively, an object in the feature space will be located in a sparsely populated region if it is an outlier, and it will be located in a dense region if it is an inlier. We use $k$NN to detect outliers using two different methods:

\begin{itemize}
\item {\bf Mean Distance to Neighbors: } We search the database and derive the $k$ nearest neighbors for each object. For each of the $k$ neighbors found, we compute each object's mean distance to its neighbors. We assign each object an outlier ranking (described above) where the object with the greatest distance to its $k$ neighbors is assigned $r_i=1$.
\item {\bf Reverse $k$ Nearest Neighbors: } Consider that objects that are frequently in the set of $k$NN for many other objects are likely to be inliers, whereas those objects that are rarely found in the set of $k$NN for other objects are outliers. This is computed using the reverse $k$NN (R$k$NN), where a query feature vector finds all of the instances where it is found in the $k$NN set of all of the other feature vectors in the database~\citep{tao2007multidimensional}. This is also known as the in-degree of a $k$NN graph. Each object is assigned a ranking described above, where $r_i=1$ denotes the object with the smallest in-degree in the dataset.
\end{itemize}

In addition to using $k$NN, we also use distance similarity searches to derive outlier scores. This algorithm simply searches a fixed radius, $\epsilon$, and returns those neighbors that are found within the search radius~\citep{GOWANLOCK2019}. The object with the fewest number of neighbors within $\epsilon$ has the greatest outlier score, $r_i=1$, and the object with the greatest number of neighbors within $\epsilon$ is the most typical object in the database.

While the $k$NN and distance similarity search methods output an outlier ranking, we also plan to use other methods to detect outliers, including those that return a binary classification (inlier or outlier). An example is density-based clustering~\citep{Gowanlock2019b}, where inliers are assigned to clusters, and outliers are assigned to a set of noise points.

Performance is critical for outlier detection tasks, as all-pairs searches have a worst-case quadratic time complexity. We are leveraging our prior work in this area to perform fast $k$NN searches, distance similarity searches, and clustering, which has been parallelized using GPUs, and has been published in the computer science literature~\citep{GOWANLOCK2019, Gowanlock2019b, Gallet2019, GOWANLOCK2021}.

\subsubsection{Element (d): Disseminate results}

We attach data from external databases to the ZTF derived data catalog to create a value-added product. At present we include data from the following static databases:
2MASS (JHK~photometry of around 20,000~asteroids \citep{2mass}) and
SDSS (ugriz photometry of around 40,000~asteroids \citep{sdss}).
We also include data from two dynamic (that is, continually or occasionally updated) databases:
WISE/NEOWISE (two or four band thermal infrared photometry and derived diameter and albedo for around 100,000~asteroids \citep{neowise});
and
the Minor Planet Center (orbital elements and absolute magnitudes for all known Solar System objects\footnote{\url{https://www.minorplanetcenter.net/data}}).
We have also ingested solutions\footnote{\url{https://www.minorplanet.info/php/lcdb.php}} from the (asteroid) Lightcurve Database
\citep{lcdb}, 
though this compendium is not maintained and updated as uniformly as the other catalogs listed above.

To disseminate our results we will deploy a web interface as well as an API; these elements are not fully mature as of this writing. 
The final development step, which will begin later this year, will be to broadcast alerts for objects that are identified as outliers. These alerts can be used by us and others to carry out telescopic confirmation and follow-up of these objects of interest.

\subsection{Requirements and current performance}

Two timing requirements are imposed by the cadence of the sky surveys that are providing data to us: (1) Our data ingest and real-time processing cannot fall behind, which implies processing on average around
100~(ZTF) and 1000~(LSST) asteroids in the 30~second exposure times used by both surveys; and
(2) Our population outlier detection scheme cannot fall behind, which means that all computations that are carried out during the following day must be completed between sunrise and sunset at the observatory.


For the first constraint,
there are two key steps that must be carried out within a 30~second window: ingesting the data and identifying real-time outliers. 
Ingesting the data is fast ($<$1~second), and our present real-time outlier detection implementations --- where we simply query observed properties such as elongation of the measured source --- 
are also fast ($<$1~second). As we develop more sophisticated multi-dimensional outlier detection schemes we will ensure that the computational performance meets the requirements.

For the second constraint, we must compare the processing time for the entire database --- or at least the subset that has changed since the previous day --- to the available time, which could be as little as 10~hours during winter observing.
For 10,000~objects, calculating all light curve related properties (color, period, amplitude) takes about 83~CPU-hours on a single AMD EPYC 7542 CPU core. 
The average number of asteroids observed in a full ZTF night may be about 10~times this amount, and LSST another factor of 10~greater.
This computational need would take more than one day, if just a single CPU were available.
Thus, our SNAPS pipelines run on NAU's High Performance Computer (Monsoon) with multiple CPU/GPU nodes in order to complete the processing within a day.
Almost all the 83~CPU hour effort is dedicated to the Lomb-Scargle periodogram processing. To address this computational load, we have created a GPU Lomb-Scargle implementation \citep{GOWANLOCK2021} that is described in Section~6.2.1. 
With this GPU approach, we can readily complete the daytime processing within one day, using our project-specific node on Monsoon that has 2~AMD EPYC 7542 CPUs (32 cores each) and 4~Nvidia A100 GPUs.


\section{SNAPShot1: Our first data release \label{sec:snapshot1}}

Since late 2018 we have been ingesting ZTF moving object alerts in
real time, every night.
We ingest data from a Kafka data stream provided by ANTARES; this is a
substream re-broadcast of Solar System objects that appear in the main
ZTF stream. (As described above, we have also retroactively ingested alerts for objects with low real/bogus scores.)
As of this writing, our database contains some 17~million
observations of some 469,000~unique asteroids. This Oracle database of measurements is 3.9~GiB as of this writing\footnote{Note that 1~GiB is $2^{30}$~bytes, whereas 1~GB is $10^9$~bytes.
Throughout the paper, when reporting data sizes, we use GiB, which uses base~2 and is how computers measure storage (and which is also the standard unit in the computer science literature). Since 1~GB is equal to $\sim$0.93~GB, they are nearly equivalent, 
and a reader can acceptably interpret GiB as GB without misunderstanding our results.}.

\begin{figure}[t]
\begin{center}
\includegraphics[width=0.45\textwidth]{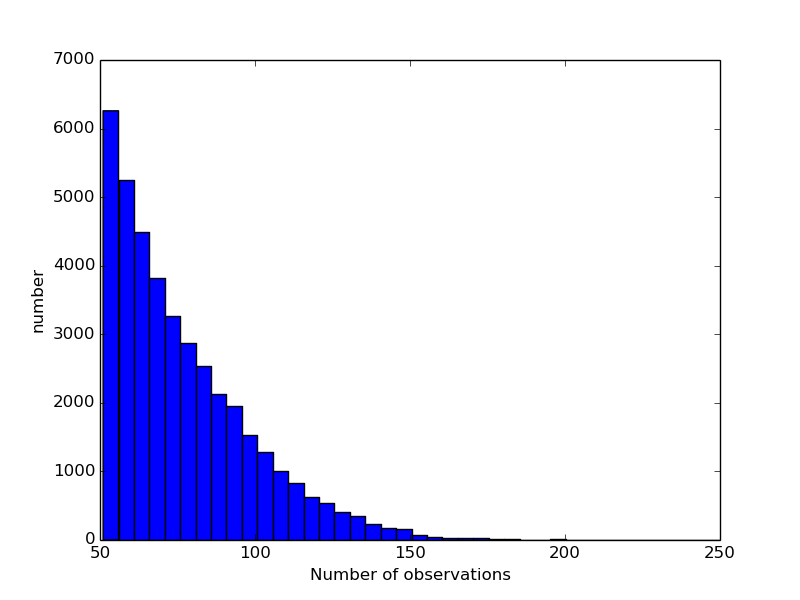}
\end{center}
\vspace{-3ex}
\caption{Histogram of number of observations of asteroids in SNAPShot1. All 40,000~asteroids presented in SNAPShot1 have more than 50~observations; nearly 6000~asteroids have more than 100~observations.
\label{fig:numobs_hist}}
\end{figure}

For the purposes of this paper we define {\em SNAPShot1}, which is our first data release. SNAPShot1 includes data for numbered, known Solar System objects reported in ZTF alerts during
2018~July~19
to 2020~May~19.
Future SNAPShots will include unnumbered objects and a broader date range.
SNAPShot1 includes 5,458,459~observations of 31,693~asteroids.
Figure~\ref{fig:numobs_hist} shows the histogram of the number of observations for targets in SNAPShot1.

\begin{figure}[t]
\begin{center}
\includegraphics[width=0.45\textwidth]{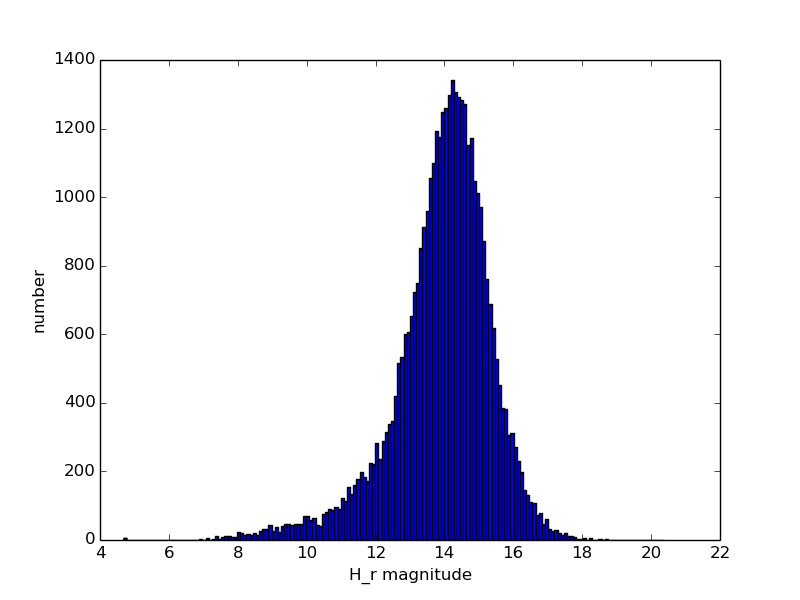}
\end{center}
\vspace{-3ex}
\caption{Histogram of derived $H_r$ magnitudes (the absolute magnitudes of the objects, in $r$ band). The histogram of $H_g$ looks very similar, with a small shift due to the $g-r$ color of each object. 
$H_r=15$ corresponds to a diameter of around 3~km (for an albedo of~0.25).
\label{fig:hr_hist}}
\end{figure}

\begin{figure}[t]
\begin{center}
\includegraphics[width=0.45\textwidth]{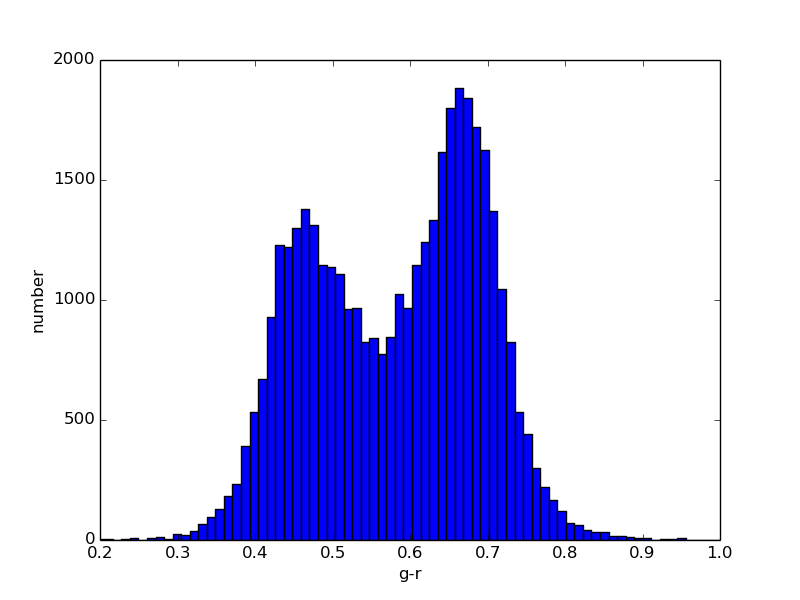}
\end{center}
\vspace{-3ex}
\caption{Histogram of derived $g-r$ colors for objects in SNAPShot1. 
  \label{fig:color_hist}}
\end{figure}

\begin{figure}[t]
\begin{center}
\includegraphics[width=0.45\textwidth]{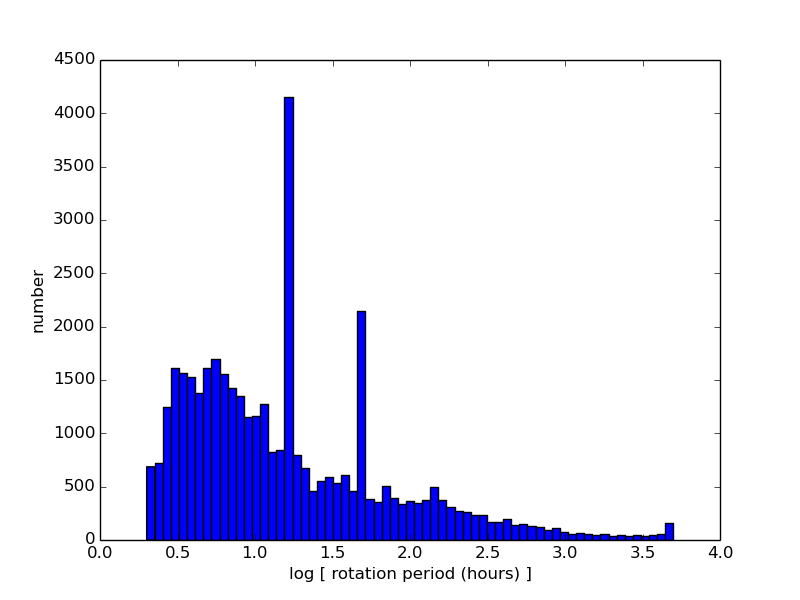}
\end{center}
\vspace{-3ex}
\caption{Histogram of derived rotation periods.
Our search range is 2--5000~hours.
This is a relatively smooth distribution except for the peaks at 16~and 48~hours, which are aliases that arise from the natural diurnal observational cadence of ZTF.
However, it is clear from interpolating across these alias-affected bins that 
some 10\% of those solutions are likely correct, but 
it is impossible to identify which objects these are.
The small peak in the very largest bin may represent objects either that have periods longer than 5000~hours, or objects with very low amplitude lightcurves where the ``best'' period simply reflects the search window.
  \label{fig:lcper_hist}}
\end{figure}

\begin{figure}[t]
\begin{center}
\includegraphics[width=0.45\textwidth]{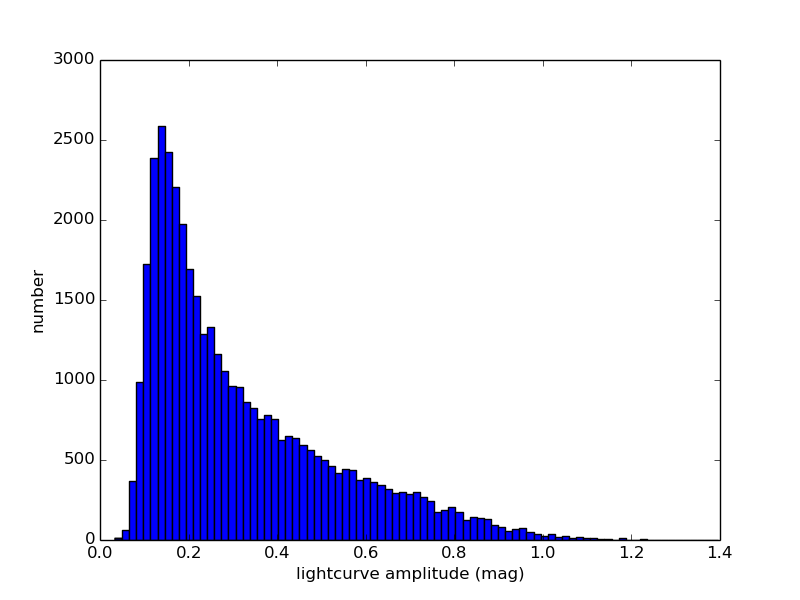}
\end{center}
\vspace{-3ex}
\caption{Histogram of derived
lightcurve amplitudes. This is a very smooth distribution, with the largest values around 1.4~mag. The peak of this histogram is around 0.15~mag, with relatively few objects having derived amplitudes smaller than this. This lack of low amplitudes is almost certainly an observational bias, as typical ZTF photometric uncertainties are in the range 1\% -- 10\% such that detecting amplitudes smaller than 0.15~mag would be difficult. 
\label{fig:lcamp_hist}}
\end{figure}

\begin{figure}[t]
\begin{center}
\includegraphics[width=0.45\textwidth]{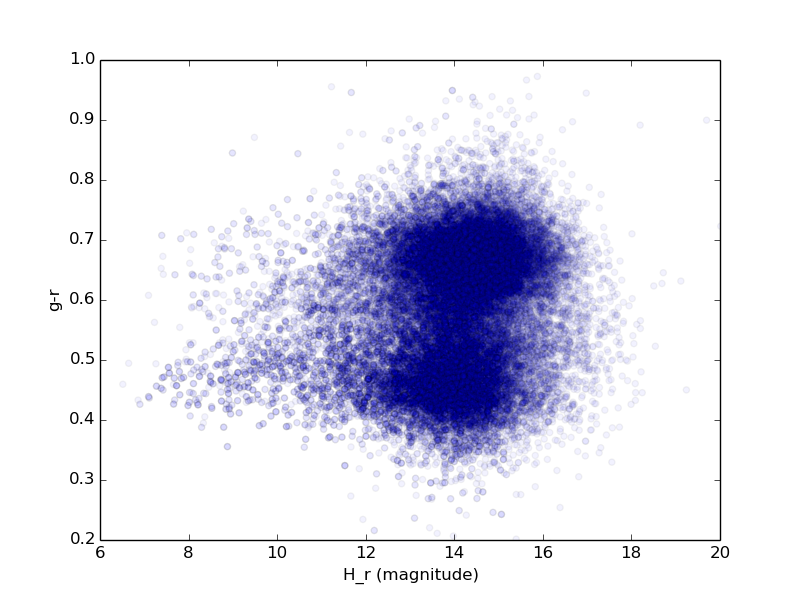}
\end{center}
\vspace{-3ex}
\caption{SNAPS $g-r$ color as a function of SNAPS $H_r$. Redder objects are detected to slightly smaller sizes (larger $H$ values) than less red objects.
This is presumably due to observational bias, probably because red objects have higher albedos 
(see Figure~\ref{fig:albedo_versus_grcolor})
where a given limiting optical band magnitude corresponds to larger $H$ (smaller asteroid).
\label{fig:grcolor_versus_h}}
\end{figure}

\begin{figure}[t]
\begin{center}
\includegraphics[width=0.45\textwidth]{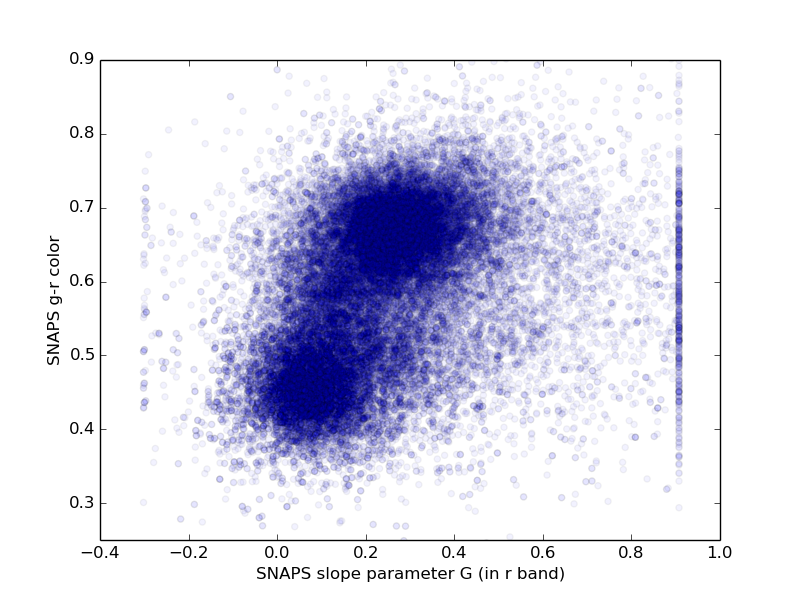}
\end{center}
\vspace{-3ex}
\caption{SNAPS $g-r$ color as a function of SNAPS slope parameter $G$ in $r$ band.
Two clear groups are shown, as expected: C-type asteroids have bluer colors and smaller slope parameters, and S-type asteroids have redder colors and larger slope parameters.
The solutions at $G=0.9$ and $G=-0.3$ indicate a small number of poor solutions rather than significant populations of asteroids with extreme slope parameter values.
Note that the commonly used default G value of~0.15 does not really correspond to an ``average'' asteroid but rather a C-type asteroid.
\label{fig:G_versus_gr}}
\end{figure}

\begin{figure}[t]
\begin{center}
\includegraphics[width=0.45\textwidth]{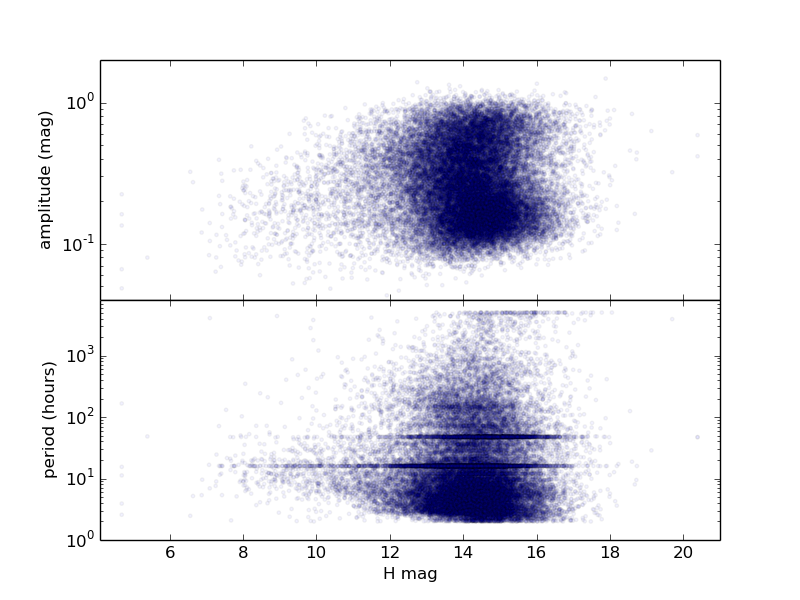}
\end{center}
\vspace{-3ex}
\caption{SNAPS lightcurve amplitude (top) and period (bottom) as a function of SNAPS H magnitudes. 
There are few large (small $H$) asteroids with amplitudes around 1~magnitude, showing that large asteroids are generally more spherical than small asteroids.
Alias period solutions are evident in the lower panel.
There are few long periods among the largest objects, but the total number of large ($H<10$) objects is small and this finding may not be significant.
\label{fig:lcamp_lcper_versus_h}}
\end{figure}

\begin{figure}[t]
\begin{center}
\includegraphics[width=0.45\textwidth]{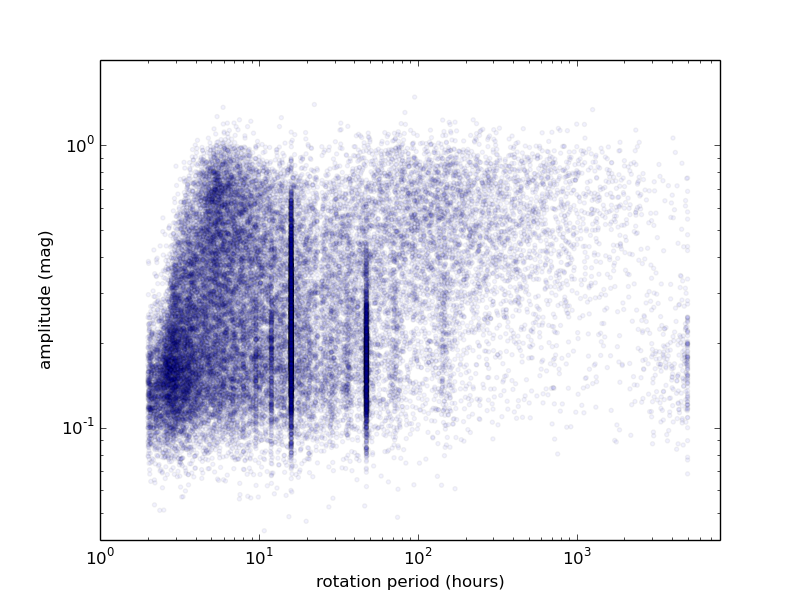}
\end{center}
\vspace{-3ex}
\caption{SNAPS amplitude as a function of SNAPS period.
Given this kind of sparse data, 
very long periods with low amplitudes would be difficult to detect, and there is a relative lack of objects in this part of the figure.
The lack of objects with short periods and high amplitudes is real and discussed further in Section~\ref{sec:strengths}.  \label{fig:lcamp_versus_lcper}}
\end{figure}

\begin{figure}[t]
\begin{center}
\includegraphics[width=0.45\textwidth]{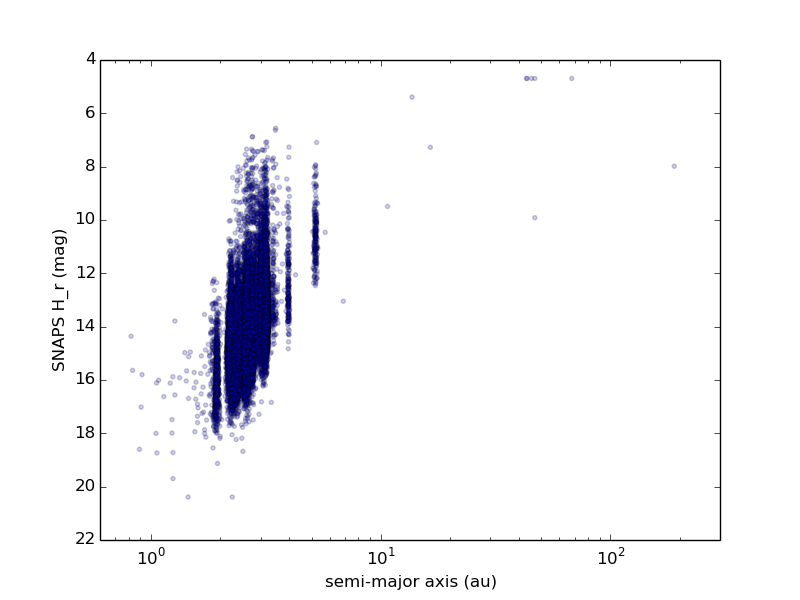}
\end{center}
\vspace{-3ex}
\caption{SNAPS $H$ magnitude as a function of semi-major axis (from the MPC). The dynamical structure of the asteroid belt is clearly evident (Kirkwood gaps, Hildas and Trojans, etc.). As expected,
ZTF is sensitive to smaller objects in the inner belt than in the outer belt and beyond: our smallest main bet asteroids have $H$$\approx$17.5 (1~km) whereas the smallest Trojans have $H$$\approx$12.5 (10~km). 
There are a small number of outer Solar System objects in SNAPShot1.
\label{fig:h_versus_a}}
\end{figure}

\begin{figure}[t]
\begin{center}
\includegraphics[width=0.45\textwidth]{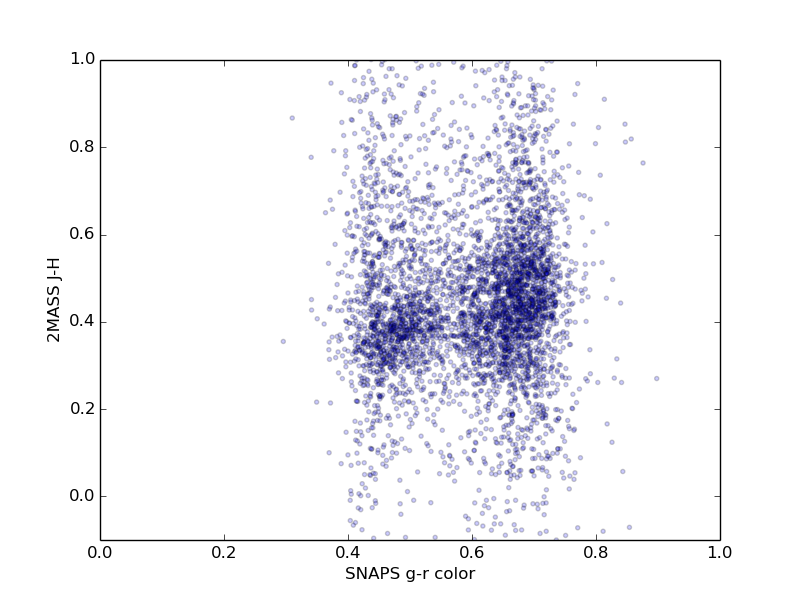}
\end{center}
\vspace{-3ex}
\caption{2MASS $J-H$ color as a function of SNAPS $g-r$ color. Despite the large scatter in the 2MASS colors, two clumps are evident, as expected: C-type asteroids are bluer in both colors and S-type asteroids are redder in both colors. Similar signatures exist for 2MASS $J-K$ and $H-K$ as a function of SNAPS $g-r$ (plots not shown here). \label{fig:2massjh_versus_grcolor}}
\end{figure}

\begin{figure}[t]
\begin{center}
\includegraphics[width=0.45\textwidth]{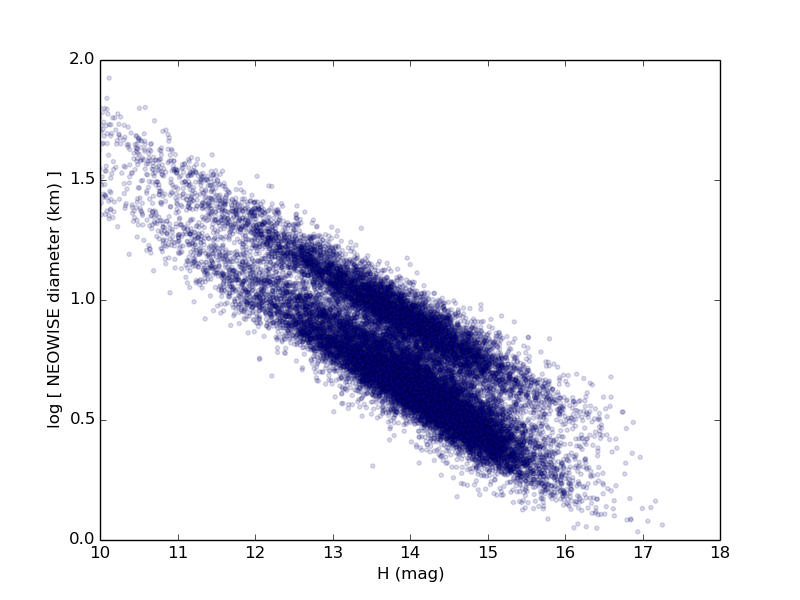}
\end{center}
\vspace{-3ex}
\caption{NEOWISE diameter as a function of derived SNAPS $H_r$ magnitude. The two parallel tracks are because S-types (lower track) and C-types (upper track) have different albedos (high and low, respectively).
\label{fig:diameter_versus_hmag}}
\end{figure}

\begin{figure}[t]
\begin{center}
\includegraphics[width=0.45\textwidth]{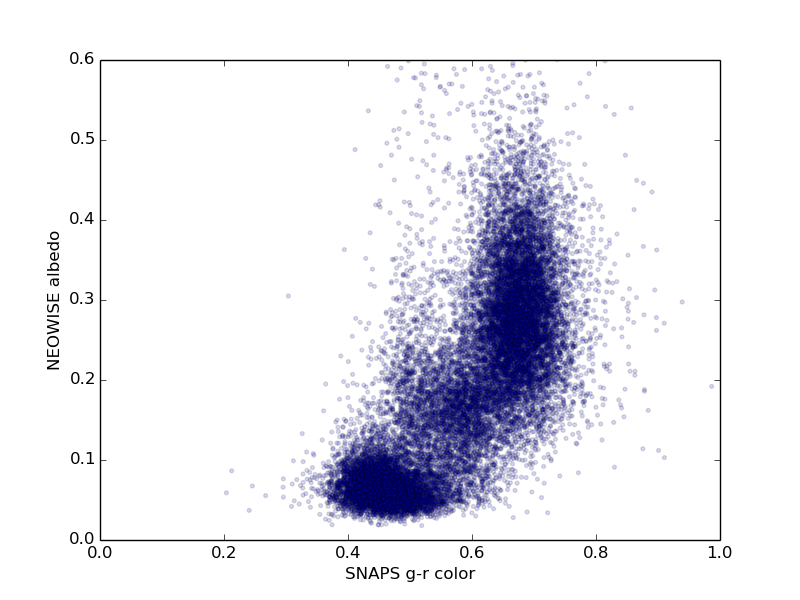}
\end{center}
\vspace{-3ex}
\caption{NEOWISE albedo as a function of SNAPS $g-r$ color. As expected, there are two broad clumps of objects present: C-type asteroids generally have less red colors and low albedos, and S-type asteroids have redder colors and higher albedos.
  \label{fig:albedo_versus_grcolor}}
\end{figure}

For each ZTF asteroid with more than 50~observations we
derive color, lightcurve period, and lightcurve amplitude, as described above, as well as absolute magnitude and
phase coefficient. 
Figures~\ref{fig:hr_hist}
through \ref{fig:lcamp_hist}
show
histograms of derived properties for objects in SNAPShot1.
Figures~\ref{fig:grcolor_versus_h} through \ref{fig:lcamp_versus_lcper}
show various combinations of SNAPS properties.
Finally, Figures~\ref{fig:h_versus_a} through \ref{fig:albedo_versus_grcolor} show SNAPS properties in relation to properties from external catalogs. Several of these figures are discussed in the following sections.

Not shown here are several parameter comparisons where nothing new is revealed.
Grossly, we find 
that lightcurve period and lightcurve amplitude do not significantly depend on $g-r$ color (i.e., asteroid taxonomy).
There is no apparent dependence of lightcure period or amplitude on semi-major axis.
And, as expected, we find that the inner belt has more objects that have redder $g-r$ colors (S-types), and the outer belt is dominated by objects that have less red $g-r$ colors (C-types).

\section{Accuracy of our derived properties \label{sec:accuracy}}

\subsection{Accuracy of derived properties -- synthetic asteorids \label{sec:synthetic}}

To determine the accuracy of our
derived physical properties
we have
created
a population of synthetic asteroids;
a detailed description of this synthetic asteroid population is given in the Appendix.
Briefly, these asteroids are randomly assigned
colors,
lightcurve periods, lightcurve amplitudes, and several other
properties from distributions that
are similar to
distributions in the literature (generally, from the Lightcurve Database \citep{lcdb}). 
These synthetic asteroids are ``observed'' --- that is, a synthetic observational
record is produced --- with ZTF-like cadences and ZTF-like
photometric errors. These datasets are then
passed through our
processing pipeline, which derives the various properties. Finally,
we compare the derived properties to the input properties to
assess the accuracy of our approaches.


Some results from analyzing this synthetic population are
shown in Figures~\ref{fig:synth_colorcolor}--\ref{fig:synth_lcperlcper}.
For our derived colors,
around 60\% of the solutions are within 10\% of the assigned colors, and 90\% of the solutions are within around 25\% of the assigned colors.
For our derived lightcurve amplitudes,
around 60\% of the solutions are within 10\% of the assigned lightcurve amplitudes, and 90\% of the solutions are within around 20\% of the assigned amplitude.
Finally, for our derived lightcurve periods,
around 70\% of the solutions are within 10\% of the assigned periods (here we do not include aliases as ``matching''), and
90\% of the derived periods are within around a factor of~2. 
Based on this analysis we conclude that in general our pipeline is producing reliable results for our derived properties. However, as with any tool that carries out bulk processing of large datasets, any individual asteroid result may have a derived property that is not correct. The SNAPS database can readily be used for bulk analysis (i.e., distributions of periods or amplitudes),
and solutions for most individual asteroids are correct.
However, individual objects that are found to be unusual in our processing (example: unusually high amplitudes, or extreme colors, or long periods) should be examined carefully.


\subsection{Accuracy of derived properties -- comparison to other published results}

\begin{figure}[t]
\begin{center}
\includegraphics[width=0.45\textwidth]{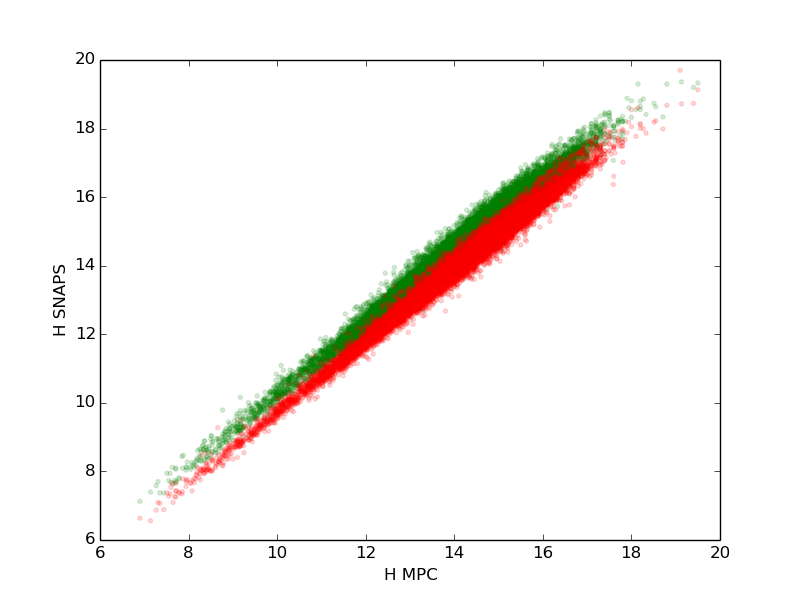}
\end{center}
\vspace{-3ex}
\caption{SNAPS H magnitudes
in $g$ and $r$
 (green and red symbols, respectively) as a function of H magnitude from the MPC, which is taken to be $H_V$.
 The offset between the green and red locus of points shows the mean $g-r$ color of asteroids. The overall agreement is excellent.
  \label{fig:h_versus_h}}
\end{figure}

We compare our SNAPS results to other published results. Figure~\ref{fig:h_versus_h} shows our derived SNAPS H magnitudes (in $g$ and $r$ bands) compared to the $H$ values (taken to be $V$ band) in the current version of MPCORB.DAT from the Minor Planet Center.
The correspondence is excellent,
with $<$$H_g/H_V$$>$~=~1.03 and 
$<$$H_r/H_V$$>$~=~0.99 where $<>$ indicates the mean, for all objects, of the ratio of the absolute magnitudes.
The standard deviation for both ratios is around 0.21~mag, presumably with components of around 0.1--0.15~mag, more or less, from each of the MPC and SNAPS catalogs -- typical uncertainties for survey measurements.
The offset between the two comparisons
($H_g$ and $H_r$ compared to $H_V$) simply reflects the mean color of all asteroids.

We compare our derived $g-r$ colors to those from SDSS MOC4 for objects that appear in both catalogs.
We find 
$<$$(g-r)_{SNAPS}/(g-r)_{SDSS}$$>$ to be~0.98, with a standard deviation of 0.12~mag, again showing excellent agreement between our results and previously published results.

\begin{figure}[t]
\begin{center}
\includegraphics[width=0.45\textwidth]{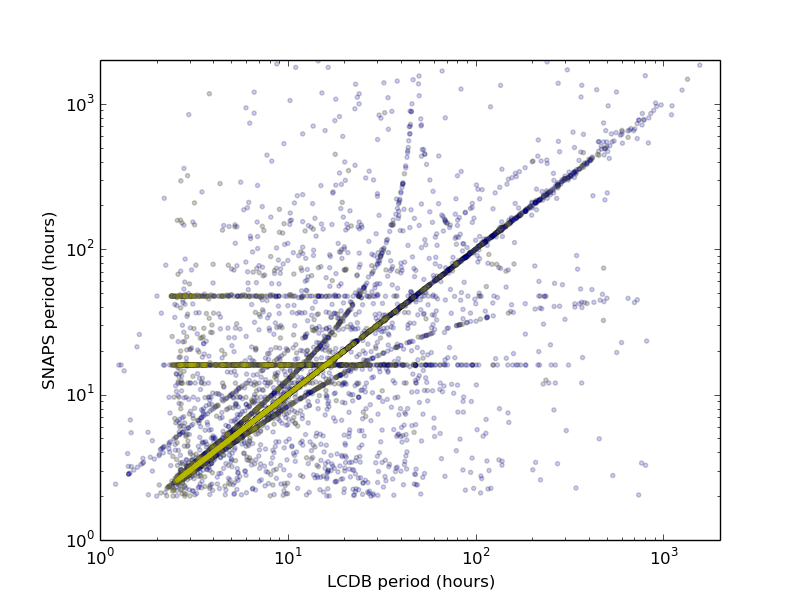}
\end{center}
\vspace{-3ex}
\caption{
SNAPS period as a function of period in the Lightcurve Database (LCDB; \citealt{lcdb}).
Blue indicates all LCDB solutions with quality code 2-~or better, whereas 
cyan shows objects with LCDB
quality codes of 3-~or better.
Around
60\% of SNAPS periods are within 10\% of the LCDB periods; when restricting the comparison dataset to LCDB objects with quality codes of 3- or better this match fraction increases to 67\%.
Most of the remaining ``unmatched'' SNAPS periods correspond to known aliases
at
16~and 48~hours (see Figure~\ref{fig:lcper_hist});
to objects where the mismatch is a factor of~2 (another common failure mode); and ``pseudo-aliases,'' which are the curved shapes in this figure \citep{vanderplas,2022FrASS...909771D}, as described in the text.
\label{fig:lcper_versus_lcdb}}
\end{figure}

We compare our derived periods to values in the current version\footnote{\url{https://www.minorplanet.info/php/lcdb.php}}
of the Lightcurve Database \citep{lcdb}. (Amplitudes cannot readily be compared since they may vary substantially from epoch to epoch due to changes in the pole orientation with respect to the observer.)
There are 8748~objects that appear in both SNAPShot1 and the LCDB.
Our period comparisons are shown in Figure~\ref{fig:lcper_versus_lcdb}.
There are several clear signatures in this plot.
(i) The 1:1 line where the two solutions agree
is readily apparent.
(ii) There are many SNAPS solutions at
periods of 16~and 48~hours (see Figure~\ref{fig:lcper_hist}). These are alias solutions that are products of the ZTF observing cadence and, in general, not the actual rotation periods for these asteroids, although some of these solutions are presumably correct (near where these alias distributions cross the 1:1 line).
(iii) There is a line parallel to the 
1:1 line, but shifted upward by a factor of~2 (in period space); this is a presumably another alias where SNAPS has derived a period that is twice that of the LCDB period.
(iv) Several curved lines can be seen that together form a quasi-triangular shape with one vertex in the lower left. These are ``pseudo-aliases'' \citep{vanderplas}
that result from interactions between the real period of an object and the strongest alias. These curves are defined by $1/P_D = \left((1/P_{R})\pm(1/\SI{24}{\hour})\right)^{-1}$ where $P_R$ is the real period for the object and $P_D$ is the derived period \citep{2022FrASS...909771D}.

Solutions in the LCDB are assigned quality codes, with ``2''~suggesting solutions that may be uncertain at the 30\% level and ``3-'' or better indicating solutions that generally are reliable. 
Almost 97\% of the
targets that appear in both SNAPShot1 and the LCDB have LCDB quality codes of 3-~or better.
We find that around 60\% of SNAPS 
periods are within 10\% of the LCDB period given a quality code of 2 or better; for LCDB code of 3- or better, the 10\% agreement rises to 67\%. 
Of the remaining solutions,
almost 20\% are in the 16~and 48~hour aliases, and most of the rest are in the pseudo-aliases.

\section{Science results to date \label{sec:science}}

The primary goal of the SNAPS project is to enable a wide range of science cases. We present here 
several science results that represent only the beginning of what can be accomplished with the large, uniform catalog that we are producing.


\subsection{Asteroid strengths \label{sec:strengths}}

Figure~\ref{fig:lcamp_versus_lcper} shows the derived rotation period and lightcurve amplitude for minor planets from SNAPS. There are two regions on this plot where the density of objects is interestingly low.
The first is the relative lack of objects showing a long rotation period with a low amplitude. This is highly likely to be an observational bias as solving for these solutions with sparse data is difficult. The second, more interesting, effect can not simply be explained by observational biases: We observe a dearth of objects at short periods ($P < 3$ h) with amplitudes larger than around 0.25~mag. This appears as an ``empty triangle'' in the upper left of Figure~\ref{fig:lcamp_versus_lcper}.

The ``spin barrier'' discussed in the context of asteroid rotation represents the critical spin rate at which a strengthless rubble pile of average density would undergo rotational fission (\citealt{pravec2000}). This effect will also be shape-dependent, as areas of the surface of an elongated body are much farther from the center of gravity of the object than would be the case for a more spherical object. It follows then that objects with shorter rotation periods
would have a more stringent limit on elongation than an objects with longer periods.


To examine this in more detail we focus here on objects with estimated diameter $0.3 < D < 20$ km to ensure that the sub-population is likely to consist of primarily rubble piles (both smaller and larger objects may be substantially monolithic, though for different reasons).
In SNAPShot1 there are 14,784~objects with
$19.5 > H > 10.5$ (where we use these $H$~magnitudes to correspond approximately to 0.3~and 20~km, respectively)
and $P < 10$~h. Using the methodology of \cite{mcneill2018}, for each object we can derive a lower limit on the required strength needed to resist rotational fission.

\begin{figure}
\begin{center}
\includegraphics[width=0.6\textwidth]{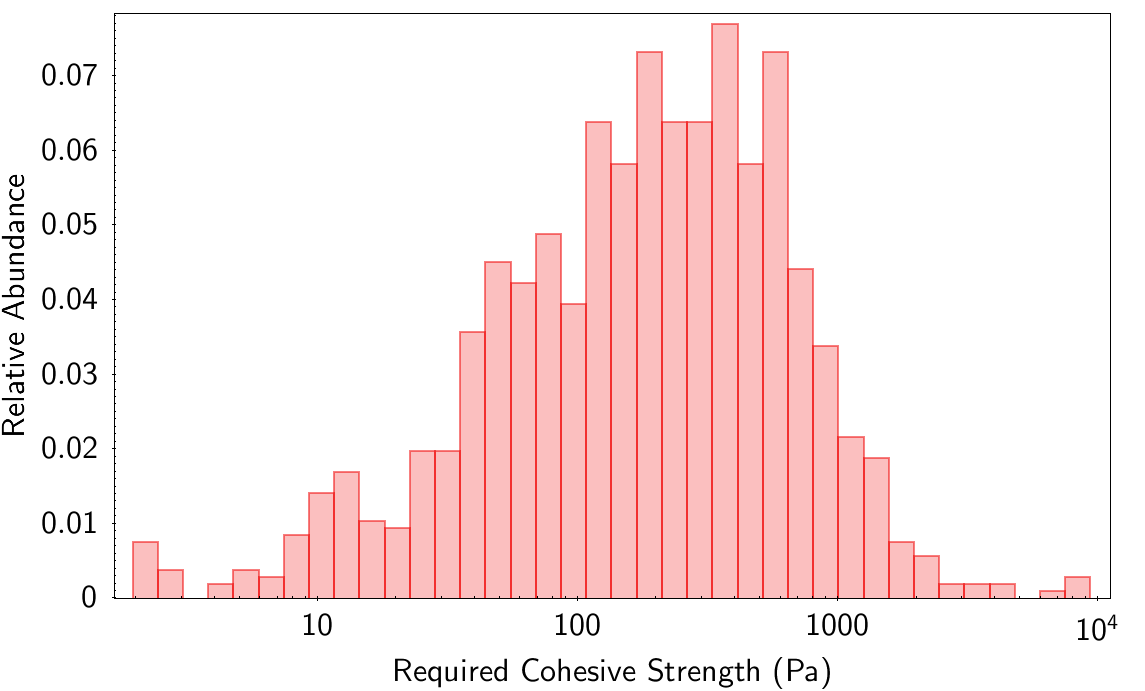}
\end{center}
\caption{Histogram showing all of the non-zero strength limits determined from SNAPS objects with $D < 20$ km, as described in the text. The majority of these objects require only minimal cohesive strength to resist rotational fission at their current shape-spin configuration, but $6\%$ require strengths greater than 1~kPa.
  \label{fig:snaps_strengthhist}}
\end{figure}

Although we can carry out the strength modelling for every object in this subset, the derived cohesive strength is always a lower limit and for most objects the value yielded is 0~Pa. As a lower limit, 0~Pa does not provide
any useful constraint, and we do not present these findings here. However, 
1067~of the 14,784~objects
in this subset
require non-zero cohesive strength if they are indeed rubble piles held together only by self-gravity and the friction between constituent parts. This is a larger sample of strengths than has been previously determined. A histogram of these non-zero strengths is given in Figure~\ref{fig:snaps_strengthhist}. Note that for this calculation all objects are assumed to have typical S-type bulk density of $2500$ kg m$^{-3}$.

\begin{figure}
\begin{center}
\includegraphics[width=0.6\textwidth]{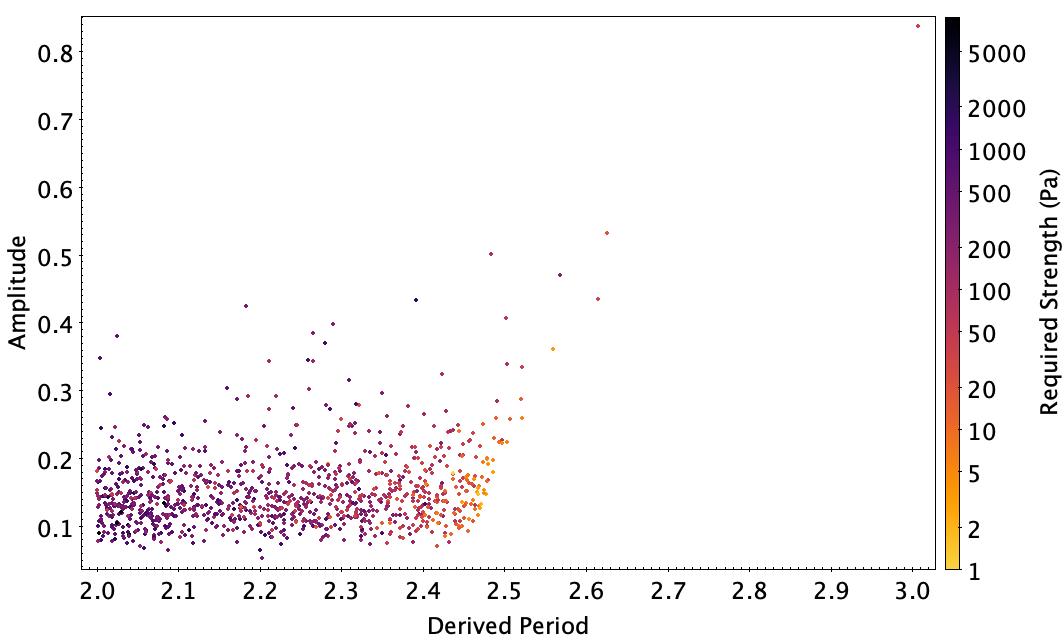}
\end{center}
\caption{The derived rotation periods and amplitudes for objects in SNAPS with derived non-zero strength limits (as shown in Figure~\ref{fig:snaps_strengthhist}). The auxiliary scale shows the required internal strength (in Pascals) for the object to resist fission.
The empty region in the lower right shows where
objects with zero strength may exist.
  \label{fig:snaps_avsp}}
\end{figure}

Figure~\ref{fig:snaps_avsp} shows the periods and amplitudes for objects shown in Figure~\ref{fig:lcamp_versus_lcper} that require non-zero strengths to resist rotational fission. The auxiliary axis shows the calculated (minimum) strength value. 
This result implies a boundary dependent on both period and amplitude beyond which 
internal cohesive strength
is required. Since this boundary depends on size and object density it will not appear as a clear cut-off in a population of mixed size and taxonomy,
as is seen in the SNAPS results in Figure~\ref{fig:lcamp_versus_lcper} and in the derived strengths of objects in Figure~\ref{fig:snaps_avsp}.

\begin{figure}
\begin{center}
\includegraphics[width=0.6\textwidth]{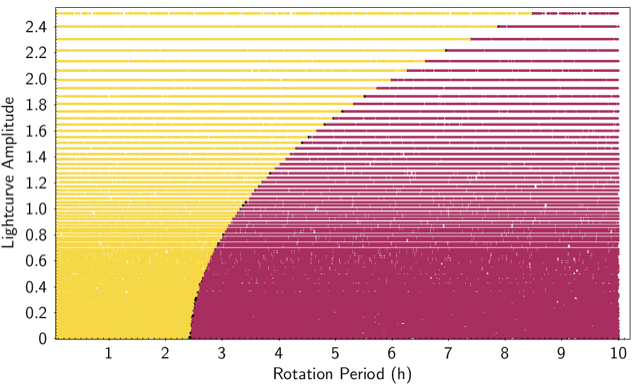}
\end{center}
\caption{The strength (yellow) and strengthless (maroon) regimes for 
a population of synthetic asteroids, each of which has a 3~hour rotation period and density of 2500~kg/m$^3$. 
  \label{fig:simulated_snaps_avsp}}
\end{figure}

We further explore this result with a synthetic population of objects having a range of periods and amplitudes, but a fixed diameter and bulk density of 3~km and 2500~kg m$^{-3}$ respectively.
Figure~\ref{fig:simulated_snaps_avsp}
shows the boundary between the strength and strengthless regimes for objects of this size and density.

The conclusion of this analysis is as follows.
Some rapidly rotating, low-amplitude SNAPS asteroids require strengths as large as 5000~Pa, but most minimum strengths are hundreds of Pascals (Figure~\ref{fig:snaps_strengthhist}). These values establish a lower limit on the strength within these bodies. The strengths required are sufficient that we must consider that there are mechanisms beyond simple friction between constituent pieces of the rubble pile preventing rotational fission. The values obtained are significantly lower than the internal strength of solid rock so we do not necessarily believe that this implies that all of these objects are monolithic in nature, as opposed to being rubble piles. Instead the strengths are similar to that of lunar regolith (\citealt{mitchell1972}) so we consider the possibility that these objects may have some lunar-like cohesion caused by regolith on their surfaces.
The lack of SNAPS objects with large lightcurve amplitudes and short periods is a real effect. Objects with these parameters --- that is, in the ``empty triangle'' of Figure~\ref{fig:lcamp_versus_lcper} --- would require significant strength to maintain those elongations and rotation periods.
The lack of objects in this area of parameter space indicates that indeed most asteroids have little or no strength.

Objects may be placed into this ``empty triangle'' corner of parameter space through (for example) collisions or the YORP effect, but unless they have sufficient internal strength they would soon shed material from their elongated ends, therefore increasing their rotation periods (through angular momentum loss) and decreasing their elongations.
This would have the effect of moving those objects downward and to the right in Figure~\ref{fig:lcamp_versus_lcper} --- out of the ``empty triangle'' --- until they cross the stability boundary implied in that figure and in Figure~\ref{fig:simulated_snaps_avsp}.
Objects that appear to require significantly large strength --- in other words, are found in or near the ``empty triangle'' --- may be good targets to search for activity that is derived from mechanical mass loss.

\subsection{Jupiter Trojan asteroids}

\cite{2021PSJ.....2....6M} used sky-survey data from ATLAS to derive the mean shape distributions of the L4 and L5 Jupiter Trojan asteroids. They conclude that L4 asteroids are, on average, more elongated than L5 asteroids, but only if their period distributions are similar; the observations could be explained without any shape difference if the L5 asteroids rotate significantly more slowly than the L4 asteroids.

\begin{figure}
\begin{center}
\includegraphics[width=0.6\textwidth]{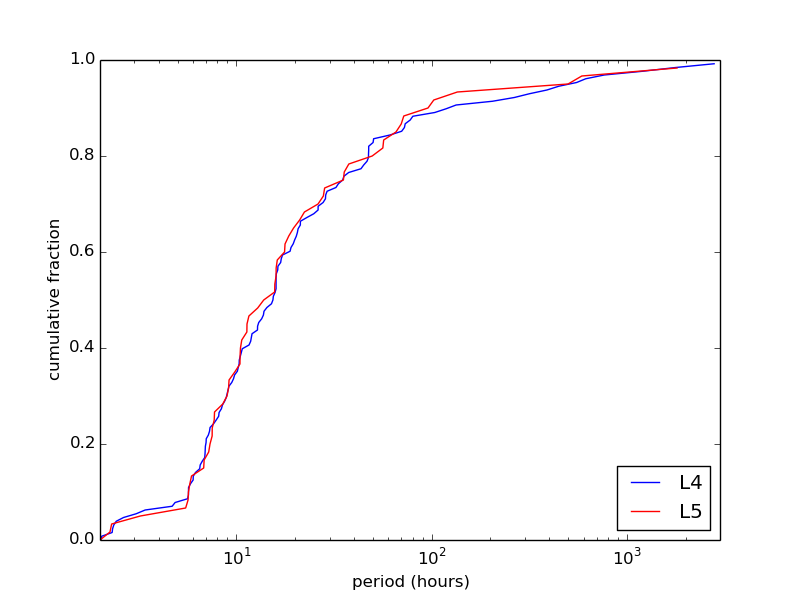}
\end{center}
\caption{Cumulative fractions for the period distributions of the L4 (blue) and L5 (red) Jupiter Trojans. The period distribution of L5 Trojans is slightly and insignificantly shifted to shorter periods than for L4 Trojans. This result rules out the possibility that L5 Trojans have significantly slower rotation rates than L4, which increases the likelihood that the collisional argument in
\cite{2021PSJ.....2....6M}
is correct.
\label{fig:jtrojans_ks}}
\end{figure}

We can compare period distributions using our SNAPS catalog.
There are 188~Jupiter Trojan asteroids that appear in SNAPShot1 (128~L4, 60~L5).
Figure~\ref{fig:jtrojans_ks}
shows the cumulative fractions for the period distributions for L4~and L5~Trojans.
We find that the L4~Trojans are very slightly slower than the L5~Trojans (blue line slightly to the right of the red line). It is clear that the L5~Trojans are {\em not} significantly slower than the L4~Trojans (which would appear as a red line significantly to the right of the blue line), and in fact
a K-S test between these two data sets indicates that the period distributions are not significantly different.
Therefore, we rule out one of the
potential alternative explanations given in \cite{2021PSJ.....2....6M}, increasing the likelihood that the two clouds do indeed have different shape distributions and that the collisional interpretation in that paper is correct.

%

\subsection{Individual objects of interest}

SNAPShot1, with more than 30,000~asteroids, contains a large number of asteroids that are not individually remarkable, but also a small number of objects that may be interesting. An exhaustive search for individual objects of interest is too cumbersome for any paper, or even any investigator; the data are available to enable a large number of people to carry out a wide range of science experiments. Here we present a few individual objects of note, to demonstrate several aspects of working with our database of derived properties.

\subsubsection{(468861) 2013 LU28}

(468861) 2013 LU28 is the object in SNAPShot1 
with the largest semi-major axis: around 180~au (Figure~\ref{fig:h_versus_a}).
This object has an orbit
that is Centaur- or comet-like, with
eccentricity~0.95 
(perihelion is 8.7~au)
and inclination~125~degrees (in other words, a retrograde orbit).
Its absolute magnitude is~8.0,
which would make it one of the largest comets known, perhaps slightly or somewhat larger than the remarkable and recently discovered comet C/2014~UN271 (Bernardinelli-Bernstein). Our contribution to the understanding of this object is our derived lightcurve properties. We derive a 
rotation period of 19.5~hours and a lightcurve amplitude of 0.19~mag, suggesting a body that may be elongated or aspherical at the 10\% -- 20\% level, which is not remarkable.
We cannot constrain the rotation pole.
This object is presently around 10~au from the Sun, a distance at which many Centaurs and comets show activity (see, for example, Table~1 in \citet{2020ApJ...892L..38C}).
Therefore, further monitoring of this object --- by classical observers, and through our SNAPS active object detection --- may soon show interesting evidence of activity, potentially shedding light on the origin of this unusual body.

\subsubsection{(1620) Geographos}

\begin{figure}
\begin{center}
\includegraphics[trim={0 0 0 5cm},clip,width=0.6\textwidth]{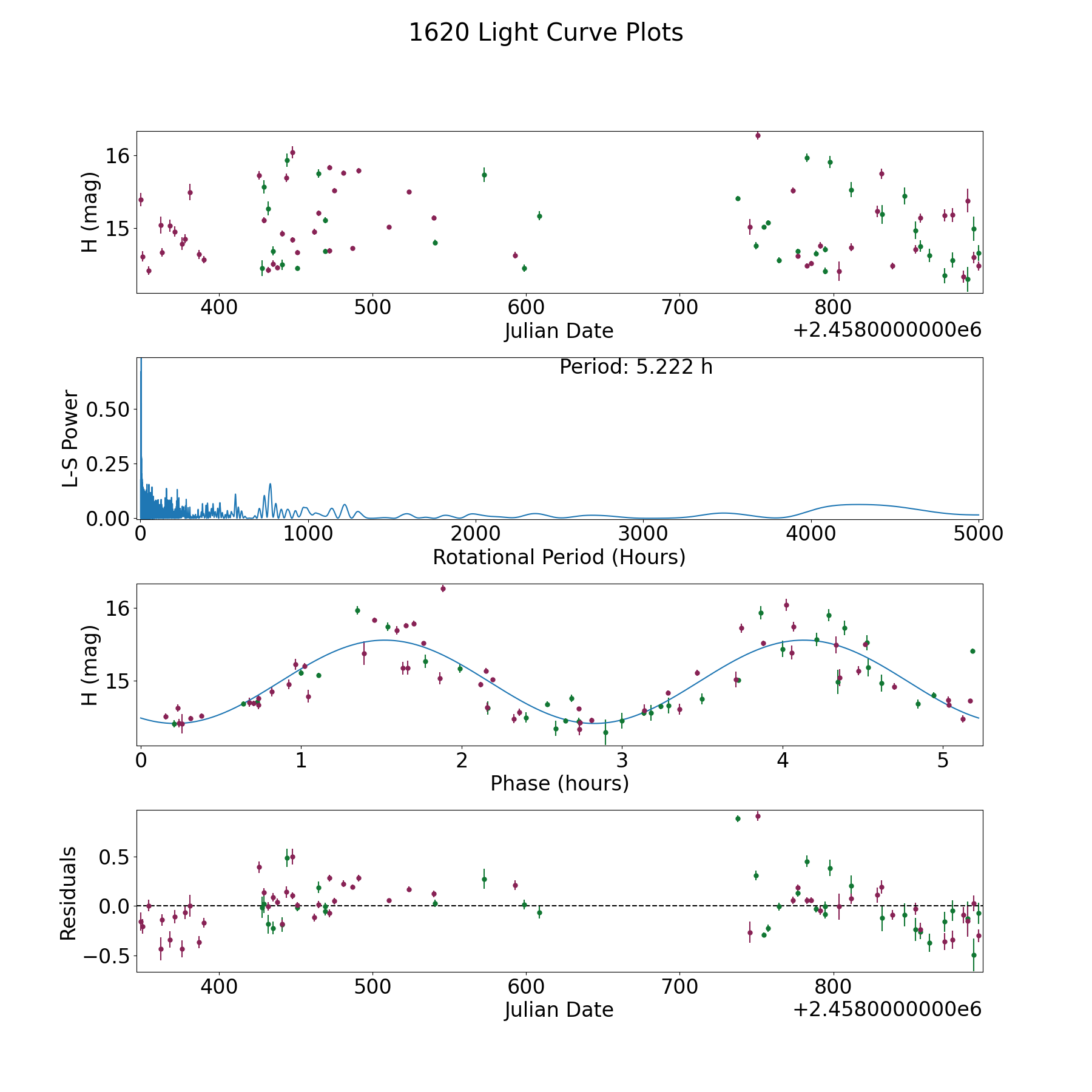}
\end{center}
\vspace{-5ex}
\caption{SNAPS results for (1620) Geographos. We produce analogous four-panel figures for each of our 31,693~targets using SNAPShot1 data. For all panels, the green points are observations in the $g$ filter and red points are in the $r$ filter; 
the $g$-band data are shifted by the derived $g-r$~color.
The top panel shows $H$~magnitude for each individual observation.
The second panel shows the Lomb-Scargle Periodogram computed over a period grid from $2$--$5000$~h;
the vertical axis is normalized Lomb-Scargle Power.
The best period (highest peak in the periodogram) is labeled.
The third panel shows the $H$~magnitudes shown in the top panel, folded to the best derived period (second panel), with a fitted sine curve. Phase (horizontal axis) is shown in hours, from zero to one full rotation.
The bottom panel shows residuals: folded data minus the fitted sine curve. 
\label{fig:1620}}
\end{figure}

(1620) Geographos 
is a Near Earth Object (NEO), one of only~95 in SNAPShot1. (NEOs generally have less regular observing seasons, so fewer of them will meet our criteria of 50~or more observations. Additionally, a relatively small percentage of NEOs are numbered, and SNAPShot1 only includes numbered objects. Future SNAPShots will have larger NEO samples.) 
This well-studied NEO has a known 
period of 5.22~hours and a reported amplitude in the range 1--2~mag in the Lightcurve Database\footnote{\url{https://www.minorplanet.info/php/lcdb.php}}.
(Amplitudes may vary from season to season depending on the relative orientation of the observer and the rotation pole.)
We find a period of 5.22~hours and an amplitude of 1.14~mag (Figure~\ref{fig:1620}), in excellent agreement with previously published values. This example is included here to show that our standard processing is acceptable even for high-amplitude NEO lightcurves, which generally present more complications than unremarkable main belt asteroids because of irregular observing seasons and wider phase angle ranges.

\subsubsection{(1865) Cerberus}

\begin{figure}
\begin{center}
\includegraphics[width=0.6\textwidth]{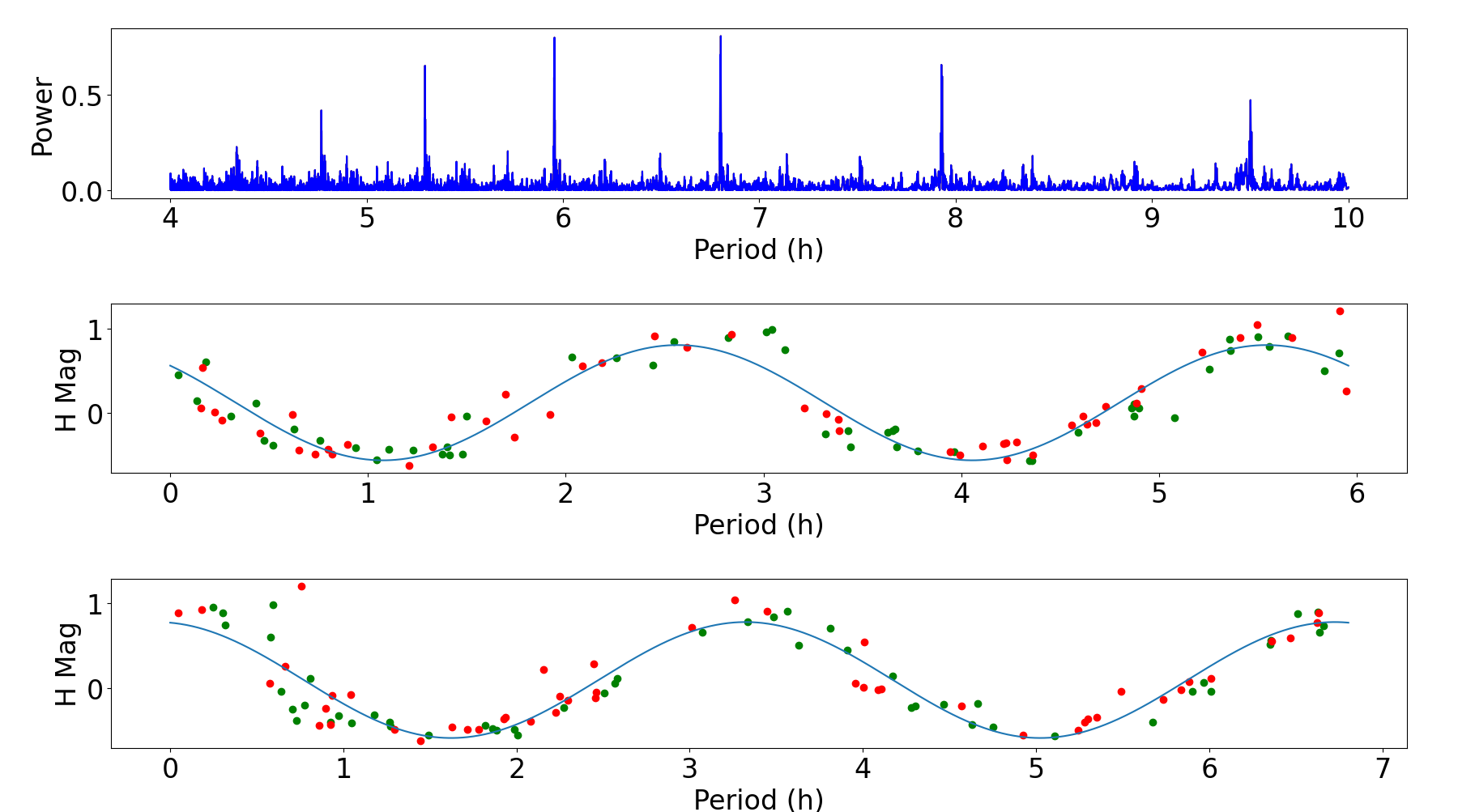}
\end{center}
\caption{SNAPS results for (1865) Cerberus.
The top panel shows our periodogram for this object in the range 4--10~hours. 
The peaks have a ``picket fence'' shape where the solutions in frequency space differ by the constant factor of (1/48~hr). This aliasing gives rise to the curved ``pseudo-aliases'' shown in Figure~\ref{fig:lcper_versus_lcdb}.
The middle panel shows the data folded to our best solution 5.95~hours, and the bottom panel shows the data folded to the literature value of 6.8~hours. Both folded solutions look acceptable.
We present this object to demonstrate that our processing may find acceptable and good solutions that nevertheless contradict published values, or other data. Some caution is therefore urged in using results for any specific object from SNAPShot1.
\label{fig:1865}}
\end{figure}

Asteroid (1865) Cerberus is also a well-studied NEO.
For this object we derive a period of 5.96~hours and an amplitude of 1.4~mag, which is well within the range of amplitudes reported in the Lightcurve Database (0.56--2.3~mag).
However, this NEO has a published period of
6.8~hours,
as indicated by many entries in the Lightcurve Database, with the earliest reported measurement being from \citealt{1989Icar...81..314H}.
The SNAPShot1 data, folded at both periods, is shown in Figure~\ref{fig:1865}.
Both folded lightcurves appear acceptable.
We present this object to demonstrate that our processing may find acceptable and good solutions that nevertheless contradict published values, or other data, and the SNAPS result may or may not be correct. Some caution is therefore urged in using results for any specific object from SNAPShot1. As demonstrated in Section~\ref{sec:accuracy},
most of our solutions are correct, but even among solutions that are credible some may still be incorrect.

\section{Discussion \label{sec:discussion}}

\subsection{Other recent SNAPS-related results}

\citet{erasmus2021}
present 
the discovery of very slowly rotating asteroids, with 39~objects each having periods greater than 1000~hours. This analysis was enabled by our SNAPS processing -- in this case, our derived lightcurve periods. That paper presents the existence of such very slowly rotating objects, and estimated that very slow rotators must be at least 0.4\% of all asteroids.
In the current database we find that 633~asteroids (around 2\%) have rotation periods greater than 1000~hours. 
We find 83~objects that have period solutions of exactly 5000~hours, our maximum allowed period, which could either indicate a true period that is longer than 5000~hours, or instead a low amplitude lightcurve; excluding these objects yields
550~very slowly rotation asteroids, or around 1.7\% of the total sample.
These very slow rotators would be extremely difficult to detect in classical lightcurve programs (one person, one telescope, one target at a time).

As an outgrowth of our SNAPS work,
\citet{navarromeza}
showed that
the derived size and shape distributions for asteroids near the detection limit of any large-scale sky survey may be biased due to non-zero lightcurve amplitudes. The catalogs produced from ZTF and LSST will need to be debiased in order to derive accurate distributions of these properties.

\subsection{Operating at LSST scale}
LSST will produce roughly an order of magnitude more data than ZTF. The two major computational tasks that need to be carried out are deriving properties of asteroids, and outlier detection to find outlying asteroids. As has been described in other ZTF processing pipelines for the detection of transient events, deriving periods is often the most computationally expensive task~\citep{coughlin2021ztf}, and we address our solutions to this problem in Section~\ref{sec:gpu_LS}. In Section~\ref{sec:db_computational resources} we briefly discuss what SNAPS will store in its database, and the computational resources available to SNAPS that will be used during the first few years of the survey.

\subsubsection{Fast Period Searches Using the Lomb-Scargle Periodogram}\label{sec:gpu_LS}


SNAPS will derive light curves of small bodies in the Solar System; however, period searches are a computationally expensive operation at LSST scale.  We examine the expected performance of SNAPS for deriving asteroid rotation periods using the LSST synthetic moving object database\footnote{\url{https://github.com/lsst-sssc/lsst-simulation}}. The database contains $\sim$43 million objects. We derived periods for a subset of objects by selecting those that have at least 50~observations in two filters (i.e., those objects with sufficient observations to produce believable light curves). This yielded a total of 63,300 objects, with an average of 517 observations per object. As expected, the maximum observing window of an object is 10 years (the length of the survey).

We examine period search performance at LSST scale using the GPU-accelerated Lomb-Scargle algorithm developed by our team~\citep{GOWANLOCK2021}. 
We derive the periods for the $\sim$63K asteroids described above, and search frequencies in the range 0.0150--150.796 day$^{-1}$, which corresponds to asteroid rotation periods of 2--20,000 hours, as LSST will enable the search for very slowly rotating asteroids. We use the method described by~\citet{richards2011machine} for selecting the frequency grid spacing, $\Delta f$, which yields roughly one million frequencies over the abovementioned frequency range.  Thus, we carry out a search using $N_f=10^6$ searched frequencies for each object on a uniformly spaced frequency grid. 

\begin{figure}[t]
\begin{center}
\includegraphics[width=0.45\textwidth]{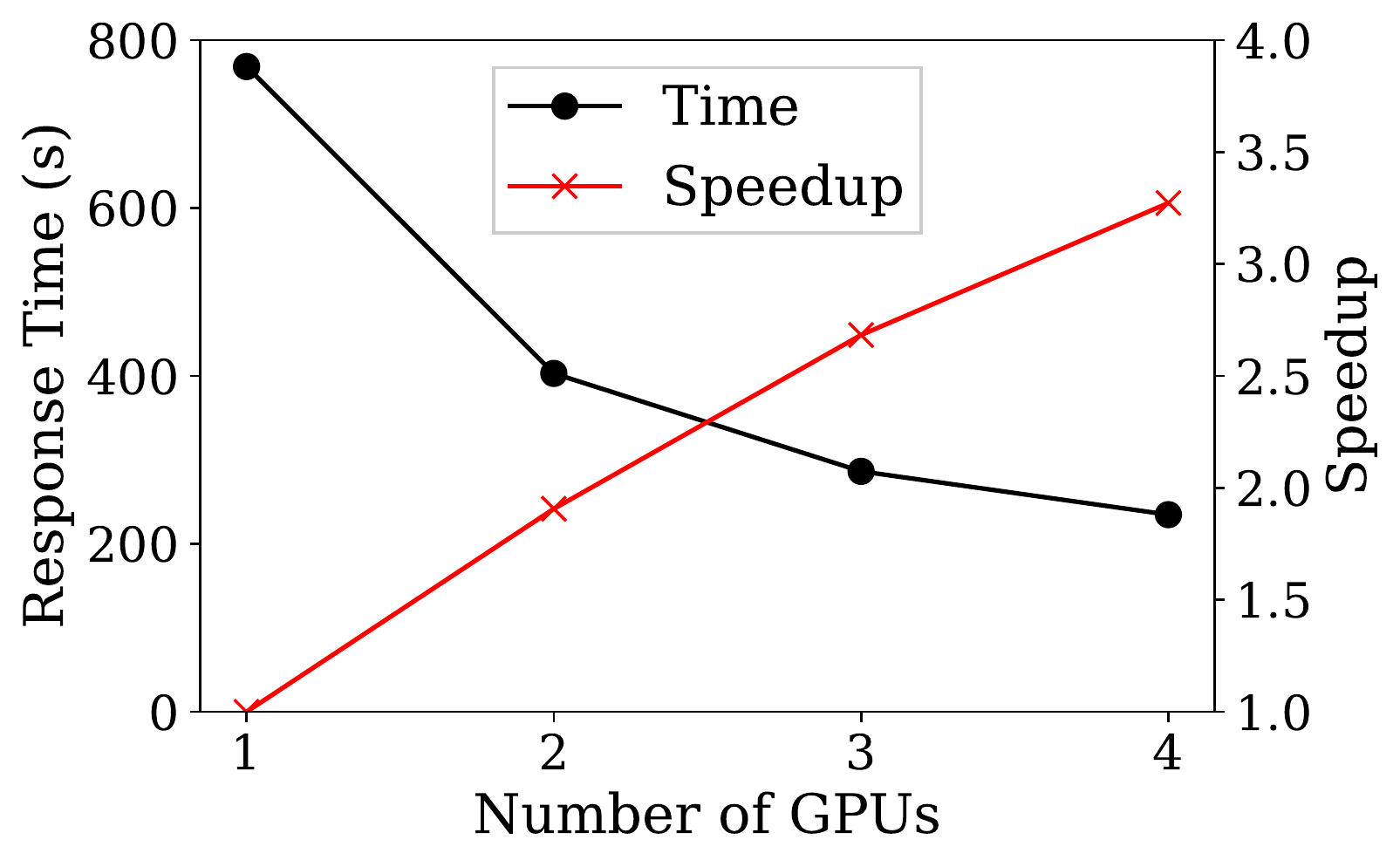}
\end{center}
\vspace{-3ex}
\caption{The response time (s) (left axis) and speedup (right axis) as a function of the number of GPUs for performing a period search using the Lomb-Scargle periodogram. We compute the periods for 63,300 asteroids, using a uniform frequency grid having $10^6$~frequencies. Our platform consists of two 2.6~GHz Intel Xeon Platinum~8358 processors with 64~total physical cores, 512~GiB of main memory, and four A100 GPUs, each with 40~GiB of on-card global memory.}
  \label{fig:LS_scalability_A100_UW_Synthetic}
\end{figure}

To assess the performance of period searches using the Lomb-Scargle periodogram, we include the time to transfer all data to the GPU, and send the periodograms back to the host for each object. The total amount of memory required to store the periodograms in 64-bit double precision format is 471.6~GiB. To ensure that the on-card global memory is not exceeded on the GPU(s), the Lomb-Scargle algorithm automatically batches the computation (see our prior work on our GPU-accelerated Super Smoother algorithm for additional detail on multi-GPU batching in \citet{GOWANLOCK2022}). Figure~\ref{fig:LS_scalability_A100_UW_Synthetic} plots the response time as a function of the number of Nvidia A100 GPUs, where the speedup is shown on the right vertical axis. We find that we can derive the periods of all 63,300 objects in 768.6~s on a single GPU, and only 234.9~s on four GPUs. The speedup when using 4 GPUs is 3.27$\times$, where we find that we do not obtain a perfect 4$\times$ speedup due to transferring periodograms back to the host. Further performance gains can be achieved if the entire periodogram does not need to be returned to the host, such as only transferring back those periods that have a significant peak in the periodogram. Consequently, this experiment shows the worst case scenario for our algorithm. 


Overall, we find that our system can derive periods very quickly, and we will be able to process a night's worth of data on the order of minutes. If we were to derive periods for all of the estimated 5.5 million asteroids that LSST will detect in 10 years (without the observation and filter thresholds described above), this result indicates that this can be computed in less than one day on a single A100 GPU. 


\subsubsection{Database and Computational Resources}\label{sec:db_computational resources}
Our database of LSST objects will be limited to moving objects, which is overall a small ($\sim$10\%) fraction of all LSST targets.
Furthermore, since ANTARES\footnote{\url{https://antares.noirlab.edu/}} and MARS\footnote{\url{https://mars.lco.global/}} will store the postage stamps for the objects, our website and API will point to those resources, such that we do not need to store and host those images. Consequently, based on the size of our ZTF database, we estimate that our database of Solar System objects in the LSST catalog will be $O(1000)$ GiB over 10 years.
Our database is hosted on an enterprise database server at NAU. 


Regarding computational resources, we have dedicated access for nighttime processing to the 4$\times$A100 GPUs that were detailed in Section~\ref{sec:gpu_LS} and that are installed in a node in NAU's compute cluster, Monsoon. Furthermore, since the node has 512 GiB of main memory, we will be able to store the entire subset of the database needed for a given night in main memory; therefore, we do not anticipate any performance-related issues regarding data movement from the database to the compute node. 

As a contingency plan, if memory resources are exhausted and performing all computation in-memory is no longer possible, we have an alternative plan to reduce the potential overhead of accessing the database. Monsoon uses a fast parallel filesystem for scratch storage. Before processing the data each night, we will stage the data from the main SNAPS database on the scratch filesystem, and then directly access that data during the night of observing. This strategy will eliminate any slow on-demand database accesses. 

\section{Future work \label{sec:future}}

Our future SNAPS-related work over the next 1--2~years 
will focus on the following topics.

\begin{enumerate}
    \item We will continue to ingest ZTF alerts for known moving objects. (Indeed, this will continue for the duration of that survey.) This process runs with little maintenance or supervision.
  \item We will continue to ingest relevant adjacent databases (e.g., the recent Solar System observations catalog that is part of Gaia DR3 \citep{gaia}), both static and dynamic.
    \item We will continue to develop our real-time analysis tools to detect (among other things) activity, eventually using a multi-dimensional machine learning approach where {\tt elong} will be one of but not the only important parameter.
    \item We will continue to develop our day-time unsupervised machine learning analysis tools to detect population outliers within our sample. Part of this development will be to ensure that we can
    operate automatically at a speed appropriate for the expected LSST data volume.
\item We will continue to develop our community access tools. At present there is an alpha version (internal testing only) of a web portal that will move soon to a beta release (testing by a small number of outside users). Furthermore, we will develop and release alpha and beta versions of an API to interact directly with our database. When mature, both the web and API interfaces will be released to the public.
\item There is now a LSST Solar System Products Database (SSPDB\footnote{\url{https://github.com/lsst-sssc/lsst-simulation}}), which is a full-scale simulation of LSST moving object observations. At present, the SSPDB does include asteroid colors but not asteroid lightcurves, and we are working with the SSPDB authors to implement realistic asteroid lightcurves. When the next version of the SSPDB, with lightcurves, is released, we will carry out the synthetic population fidelity testing described in Section~\ref{sec:synthetic} and in the Appendix at LSST scale.
\end{enumerate}

The goal of these future activities is to demonstrate our readiness to operate our high-fidelity pipeline at LSST scale in advance of the first LSST alerts in~2024.

\section{Summary \label{sec:summary}}

In this paper we present SNAPS, the Solar System Notification Alert Processing System, a broker that 
ingests data from all-sky surveys and automatically derives properties for a large number of asteroids.
We describe the architecture and demonstrate the fidelity of our derived properties. We present the current snapshot of our database --- SNAPShot1 --- and the ensemble properties of 31,693~asteroids. In the near future we will continue to ingest ZTF alerts and develop additional community access tools. We will be ready to ingest LSST alerts and work at LSST scale when that survey begins to observe and report moving objects.
 
\begin{acknowledgments}
We acknowledge many useful conversations with Colin Chandler; with Tom Matheson and the ANTARES team; with Mike Kelley, Henry Hsieh, and Davide Farnocchia; and with Mario Juric and Siegfried Eggl. This work has significantly benefited from all of their expertise.
We thank Steve Chesley for a very helpful review of this paper.

We acknowledge support from the Arizona Board of Regents' Regents Innovation Fund. Additionally, some of the computational analyses were run on Northern Arizona University’s Monsoon computing cluster, funded by Arizona’s Technology and Research Initiative Fund.
This material is also based in part upon work supported by the National Science Foundation under Grant No.\ 2042155 to MG.

ZTF is a public-private partnership, with equal support from the ZTF Partnership and from the U.S. National Science Foundation through the Mid-Scale Innovations Program (MSIP).
The ZTF partnership is a consortium of the following universities and institutions (listed in descending longitude): TANGO Consortium of Taiwan; Weizmann Institute of Sciences, Israel; Oskar Klein Center, Stockholm University, Sweden; Deutsches Elektronen-Synchrotron \& Humboldt University, Germany; Ruhr University, Germany; Institut national de physique nucl\'eaire et de physique des particules, France; University of Warwick, UK; Trinity College, Dublin, Ireland; University of Maryland, College Park, USA; Northwestern University, Evanston, USA; University of Wisconsin, Milwaukee, USA; Lawrence Livermore National Laboratory, USA; IPAC, Caltech, USA; Caltech, USA.
\end{acknowledgments}
%

\vspace{5mm}


\software{Astropy \citep{2013A&A...558A..33A,2018AJ....156..123A},  
          Cloudy \citep{2013RMxAA..49..137F}, 
          Source Extractor \citep{1996A&AS..117..393B},
          SciPy \citep{2020SciPy-NMeth}
          }




\appendix
\counterwithin{figure}{section}

\section{Synthetic Data Processing}

Here we present the discrete steps used to create the synthetic asteroid population that is used to test the fidelity of our SNAPS processing.

\subsection{Assigned asteroid properties}

The first step is to create the synthetic population of asteroids, each of which is assigned its fundamental properties, as follows. \\

\noindent (1) We create a distribution of $g-r$ colors for asteroids by summing two gaussians, one centered on $g-r=0.4$ and the other on $g-r=0.6$, to represent C and S types, respectively. For simplicity, we assume equal number of C and S type asteroids. \\

\begin{figure}[t]
\begin{center}
\includegraphics[width=0.45\textwidth]{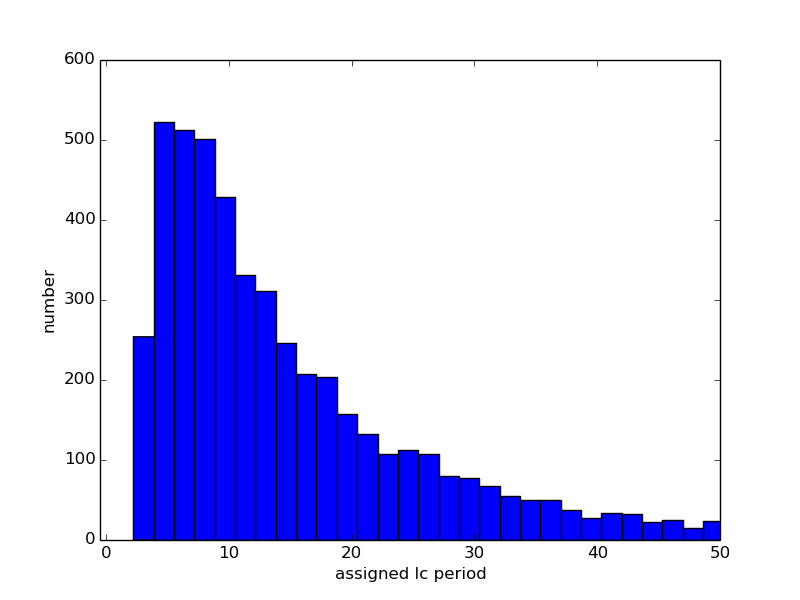}
\end{center}
\vspace{-3ex}
\caption{Histogram of assigned lightcurve periods (in hours) for our synthetic asteroids. 
  \label{fig:synth_histlcper}}
\end{figure}

\noindent (2) We assign a lightcurve period from a log-normal distribution with
\{mean, median, mode\} of 
\{19, 12, 6\}~hours. This distribution is shown in Figure~\ref{fig:synth_histlcper} and is similar to the properties of asteroid lightcurves in the LCDB. \\

\begin{figure}[t]
\begin{center}
\includegraphics[width=0.45\textwidth]{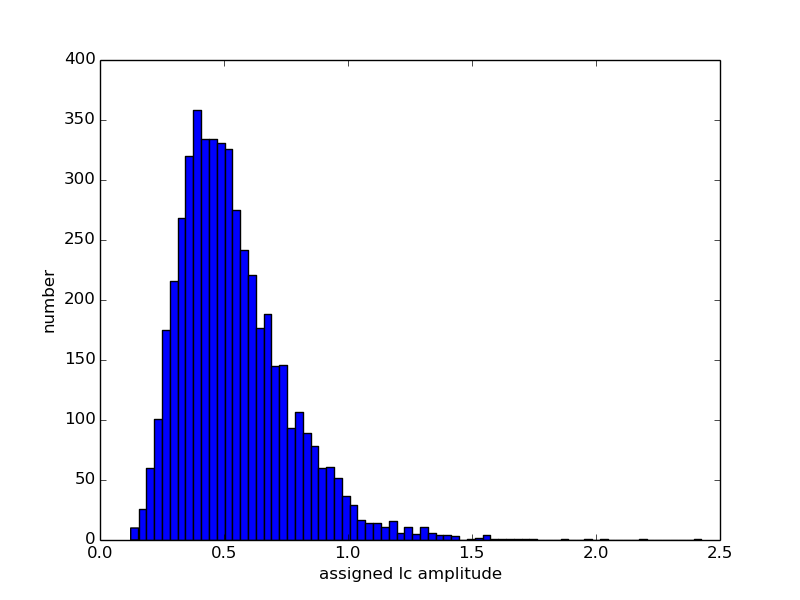}
\end{center}
\vspace{-3ex}
\caption{Histogram of assigned lightcurve amplitudes (in magnitudes) for our synthetic asteroids. 
  \label{fig:synth_histlcamp}}
\end{figure}

\noindent (3) We assign a lightcurve amplitude from a log-normal distribution with
\{mean, median, mode\} of
\{0.5, 0.5, 0.4\}~magnitudes.
This distribution is shown in Figure~\ref{fig:synth_histlcamp} and is similar to the properties of asteroid lightcurves in the LCDB. \\



\noindent (4) Finally, we assign a Solar System absolute magnitude ($H$) randomly from a power law distribution.
However, for these synthetic ZTF observations, the absolute magnitude is arbitrary, as we do not determine the detectability of our synthetic asteroids. These $H$ magnitudes may be useful in further fidelity testing for SNAPS or other projects.

\subsection{Create an observational record}

We now ``observe'' each synthetic asteroid with a relevant observing cadence. \\

\noindent (5) For each object we randomly choose a starting lightcurve phase
that will be associated with the first observational timepoint. \\

\noindent (6) We ingest observational cadences (timestamps) from actual ZTF-observed asteroids. This preserves the number of observations and aliasing generated through the actual ZTF observations. \\

\noindent (7) We ``observe'' each asteroid (that is, calculate asteroid magnitude) according to the appropriate cadence, from the previous step. For each observation we determine the magnitude as a simple sine curve, using the assigned lightcurve period and amplitude, and the initial phase assigned in step~5. \\

\noindent (8) We assign two kinds of errors for each measurement; both errors are randomly drawn from an error distribution
that is a log-normal distribution
with a mode around 0.1~mag.
This first error is applied as a random offset from the perfect sine curve.
The second error we assign is a reported photometric error (uncertainty or error bar) for each data point, which is of similar magnitude to the offset but is not identical in value. 
Assigning two similar but unique errors thus allows us to approximately capture the real ZTF uncertainty distribution. \\

\subsection{Ingest synthetic asteroids}

Each object is treated identically to real asteroids that are reported by ZTF: we derive period, amplitude, $g-r$~color, etc.

\subsection{Comparing derived properties to assigned properties}

\begin{figure}[t]
\begin{center}
\includegraphics[width=0.45\textwidth]{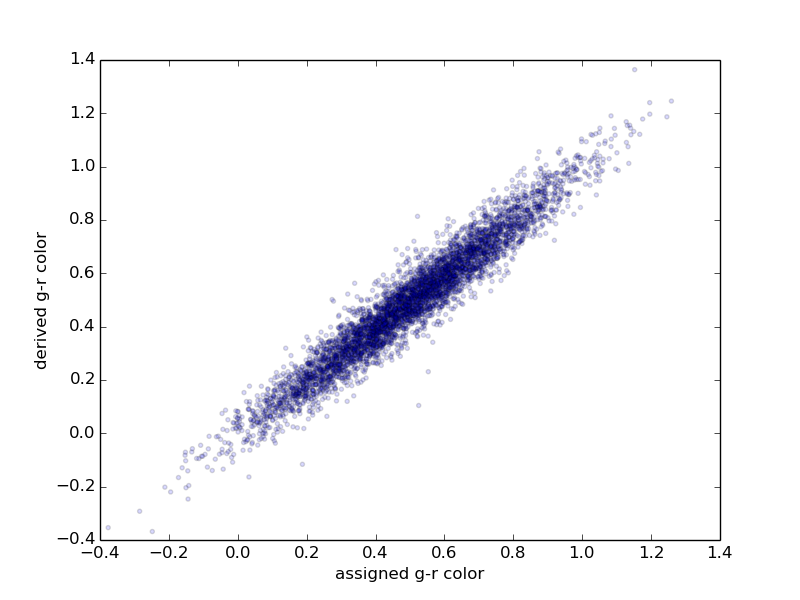}
\end{center}
\vspace{-3ex}
\caption{Derived $g-r$ color (magnitudes) versus assigned $g-r$ color (magnitudes) for our synthetic asteroids, showing that our solutions in general are good.
  \label{fig:synth_colorcolor}}
\end{figure}

\begin{figure}[t]
\begin{center}
\includegraphics[width=0.45\textwidth]{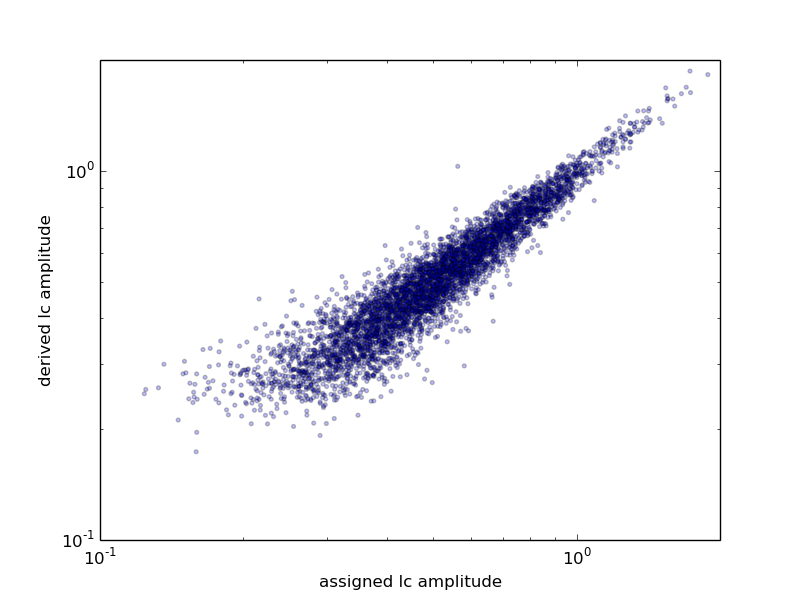}
\end{center}
\vspace{-3ex}
\caption{Derived lightcurve amplitude (magnitudes) versus assigned lightcurve amplitude (magnitudes) for our synthetic asteroids, showing that our solutions in general are good.
  \label{fig:synth_lcamplcamp}}
\end{figure}

\begin{figure}[t]
\begin{center}
\includegraphics[width=0.45\textwidth]{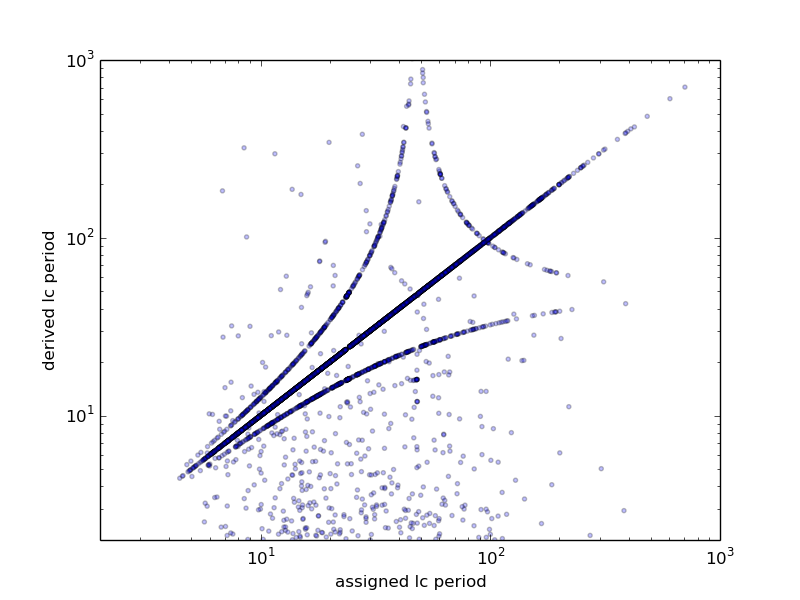}
\end{center}
\vspace{-3ex}
\caption{Derived lightcurve period (hours) versus assigned lightcurve period (hours) for our synthetic asteroids.
Most (around~70\%) points are on the 1:1~line. The curved lines are ``pseudo-aliases'' identified by \cite{vanderplas} where the ratios between the assigned and derived periods obey certain relationships as identified by \citet{2022FrASS...909771D}.
  \label{fig:synth_lcperlcper}}
\end{figure}

For each synthetic object, we compare our derived properties to the assigned properties. Results are shown in Figures~\ref{fig:synth_colorcolor}--\ref{fig:synth_lcperlcper}.
For our derived colors,
around 60\% of the solutions are within 10\% of the assigned colors, and 90\% of the solutions are within around 25\% of the assigned colors.
For our derived lightcurve amplitudes,
around 60\% of the solutions are within 10\% of the assigned lightcurve amplitudes, and 90\% of the solutions are within around 20\% of the assigned amplitude.
Finally, for our derived lightcurve periods,
around 70\% of the solutions are within 10\% of the assigned periods (here we do not include aliases such as a 2:1~ratio as ``matching''), and
90\% of the derived periods are within around a factor of~2. 

\subsection{Conclusion}

Based on this analysis we conclude that our pipeline is producing reliable results for our derived properties. However, as with any tool that carries out bulk processing of large datasets, any individual asteroid may have a derived property that is not correct. The SNAPS database can readily be used for bulk analysis (i.e., distributions of periods or amplitudes), and solutions for most individual asteroids are correct.
However, individual objects that are found to be unusual in our processing (example: unusually high amplitudes, extreme colors, or long periods) should be examined carefully.

\bibliography{snaps}{}
\bibliographystyle{aasjournal}



\end{document}